\documentclass[Journal]{IEEEtran}
\usepackage{amssymb, amsmath, graphicx, cite, epsfig, booktabs, array}

\DeclareRobustCommand{\1}[1]{\ensuremath \mathbf{1}_{\{#1\}}}

\newcommand{\naturals}{\mathbb{N}}
\newcommand{\Fig}  {\mbox{Fig.} }

\newcommand{\iid}   {\mbox{i.i.d.} }
\newcommand{\eqdef}{ := }
\newcommand{\eqdist}{ =^d }

\def\dotsim{\:\dot{\sim}\:}
\newcommand{\new}[1]{\noindent\hspace{2em}{\itshape #1 }}

\DeclareMathOperator*{\plim}{plim}

\allowdisplaybreaks[1]

\begin{document}

\title{Asymptotic Scheduling Gains in Point-to-Multipoint Cognitive Networks}
\author{Nadia~Jamal,~\IEEEmembership{Student Member,~IEEE,}
        Hamidreza~Ebrahimzadeh Saffar,~\IEEEmembership{Student Member,~IEEE,}
        and~Patrick~Mitran,~\IEEEmembership{Member,~IEEE}
\thanks{The authors are with the Department of Electrical and Computer Engineering at the University of Waterloo, Ontario, Canada (nadia, hamid, pmitran@ece.uwaterloo.ca).

 These results were presented in part at the 2009 IEEE International Symposium on Information Theory (ISIT 09).}

\vspace{-0.75cm}}
\maketitle


\begin{abstract}

We consider collocated primary and secondary networks that have simultaneous access to the same frequency bands. Particularly, we examine three different levels at
which primary and secondary networks may coexist: pure interference, asymmetric co-existence, and symmetric co-existence. At the asymmetric co-existence level, the secondary network selectively deactivates its users based on knowledge of the interference and channel gains, whereas at the symmetric level, the primary network also schedules its users in the same way.

 Our aim is to derive optimal sum-rates (i.e., throughputs) of both networks at each co-existence level as the number of users grows asymptotically and evaluate how the sum-rates scale with network size. In order to find the asymptotic throughput results, we derive a key lemma on extreme order statistics and a proposition on the sum of lower order statistics.

As a baseline comparison, we calculate the sum-rates for channel sharing via time-division (TD). We compare the asymptotic secondary sum-rate in TD with that under simultaneous transmission, while ensuring the primary network maintains the same throughput in both cases.
 The results indicate that simultaneous transmission at both asymmetric and symmetric co-existence levels can outperform TD. Furthermore, this enhancement is achievable when uplink activation or deactivation of users is based only on the interference gains
to the opposite network and not on a network's own channel gains.

\end{abstract}


\section{Introduction}

\label{sec:Introduction}
Fixed spectrum allocation policy serves as a traditional way to allow co-existence of various wireless applications in the same spectrum. Using this policy, significant portions of the limited frequency
spectrum have been assigned to licensed (primary) systems on a long term basis, while leaving only some tight bands available for unlicensed (secondary) applications \cite{FCC1}.

As the popularity of wireless services increase
and innovative systems appear, the demand for higher bandwidths and data rates grows dramatically.
On the other hand, measurements conducted by government agencies, such as the Federal Communications Commission (FCC), indicate that licensed bands are underutilized  by their licensees both temporally and geographically \cite{FCC}.
Motivated by this observation, there has been recent interest in opportunistic access of the secondary system to the licensed bands as first envisioned by Mitola \cite{Mitola}.

The basic idea of opportunistic spectrum access is to allow secondary users to communicate over licensed spectrum in a controlled fashion so that the performance of the primary system does not severely degrade.
A user with the mentioned capabilities is broadly referred to as a cognitive radio (CR).

Thus far, three different approaches namely, {\em interweave}, {\em overlay}, and {\em underlay} have been proposed to enable spectrum sharing of primary and secondary systems and consequently increase the spectral efficiency of licensed bands \cite{Goldsmith}.

In the interweave approach, the secondary system has advanced capabilities to dynamically monitor the licensed spectrum, detect frequency gaps (whitespaces), and opportunistically communicate over these gaps.
Much of the research on sequential detection of whitespaces \cite{Chen}, \cite{Lai} is
motivated by the interweave approach. Clearly, a technical challenge in this approach is developing accurate sensing methods in order to detect the presence of the primary's signal. Among methods proposed thus far, matched filter detection, energy detection,
cyclostationary detection, and wavelet detection are most common \cite{Zhang}, \cite{detection}.
In cases with multiple secondary users, cooperative detection of whitespaces helps improve performance significantly. The interested reader is referred to  \cite{Zhang} and \cite{coop1}--{\hspace{-0.1mm}}\cite{coop3} for an overview of cooperative spectrum sensing in interweave cognitive systems. After detecting whitespaces, the secondary system must apply flexible modulation schemes such as orthogonal frequency division multiplexing (OFDM) to communicate over the licensed frequency gaps \cite{Weiss}.

The overlay approach is based on the assumption that the secondary system has knowledge of the primary system's messages or possibly its codebooks. Thus, it can dedicate a part of its power to relay the primary's message and use the remaining power for
its own communication. This power split can be adjusted so that the degradation in the primary's performance caused by the interference from the secondary system is compensated for by the primary's performance enhancement due to the assistance from the secondary system \cite{Goldsmith}. Therefore, the primary system's rate can remain unchanged. On the other hand, in order to mitigate the interference at the secondary receiver and provide positive secondary throughput, the secondary transmitter can apply dirty paper coding (DPC) \cite{Costa}, based on the knowledge of the primary's messages.

The simplest overlay network is a two-user interference channel, where one user has knowledge of the other user's transmission. These systems and their capacity results are further analyzed in {\cite{Maric}}--{\hspace{-0.1mm}}{\cite{Interference2}}.

In the underlay approach, primary and secondary systems are allowed to simultaneously communicate over the same frequency bands. In this case, unlike the overlay approach, none of the primary and secondary systems have knowledge of the other system's messages or codebooks. Thus, no DPC is applied and interference is treated as noise.
Due to the constraint on the interference power at the primary receiver, the capacity of the secondary system in the underlay approach can be derived under received power constraints at the primary receiver rather than transmitted power constraints at the secondary transmitter \cite{Gastpar}, \cite{Sousa}.

In point-to-point AWGN channels, due to the non-fading characteristic of the channel, the capacity under the received power constraint and that under the transmitted power constraint are identical.
On the other hand, in fading environments, the secondary system can take advantage of the fading characteristic of the channel by
 transmitting at higher power levels whenever the channel between the secondary transmitter and the primary receiver is in a deep fade.
 In \cite{Sousa}, a specific system model is examined and the capacity of the secondary system is derived under received power constraints at the
 primary receiver. Interestingly, it is shown that an increase in the variance of the fading distribution gives rise to a higher capacity and in some cases, this capacity result is significantly more than that under a transmitted power constraint at the secondary transmitter.

In this paper, we consider collocated primary and secondary point-to-multipoint networks that simultaneously operate in the same frequency bands in a fading environment. We treat the secondary network's interference to the primary system as noise and do not allow more advanced techniques such as DPC. Since each network may operate in either of uplink or downlink modes, there are four possible uplink-downlink scenarios to be considered.

When a network operates in downlink mode, we assume that its base station transmits to the user with the highest channel gain. For the uplink transmission, we consider three different levels of co-existence based on utilization of the channel side information (CSI), i.e., the networks can apply the information on the channel and interference power gains to selectively activate or deactivate their users such that they can achieve a higher sum-rate and impose less interference to the opposite network's receiver. The levels of co-existence are further discussed in Section \ref{sec:Preleminaries}.

In \cite{Etkin}, co-existence of a large collection of
networks is considered but without the co-existence features considered in this work.

 Our main interest is studying the tradeoffs between the sum-rate of each network and the number of users. Consequently, we measure the asymptotic sum-rate of the secondary network for each co-existence level while ensuring the primary network's sum-rate is not reduced by more than a specified primary protection factor $0<f \leq 1$.
As a base reference, we consider sum-rate results of channel sharing via time-division (TD) where the primary network employs the channel for a fraction $f$ of time, leaving a fraction $1-f$ for the secondary network's transmission. Then, in each of the uplink-downlink scenarios, we compare the asymptotic secondary sum-rate under simultaneous transmission to that in TD for the same value of $f$.
 Sum-rates of primary and secondary networks in TD are similar to those in channel sharing by means of orthogonal subchannels and those in opportunistic channel sharing. In the first case, $f$ refers to the fraction of frequency bands dedicated to the primary's transmission, and, in the second, it is the fraction of licensed frequency bands in which the primary is active.

As an extension of the previous results in \cite{Jamal}, in this work, we assume channel and interference power gains to have any arbitrary CDF, possessing two specific properties. The first property corresponds to the order of the CDF in the very low gain regime while the second one corresponds to its behavior in the very high gain regime. These properties are presented in detail in section \ref{sec:Preleminaries}.
 As we shall illustrate later, the CDF of the power gains corresponding to the well-known fading distributions such as Rayleigh, Rician, and Nakagami-m have both of the mentioned properties. Therefore, the results presented in this paper can be applied to these three fading distributions.

Unlike \cite{Jamal} which considers activating users in uplink mode only based on interference knowledge, in this paper, we also consider scheduling
 users jointly based on interference to the other network and channel gains to one's own network.

The results obtained in this paper indicate that, significant asymptotic performance improvements can be gained over
TD by selectively deactivating users in uplink mode. Specifically, some conclusions that we draw are that at suitable co-existence levels

 \begin{itemize}
 \item If both networks operate in uplink mode and gains are Rayleigh or Rician distributed, in some specific cases, for $1/2 < f \leq 1$ underlay operation results in a better asymptotic sum-rate for the secondary network over TD and in some other
 cases, for $0<f<1/2$, underlay operation outperforms TD.

  \item If the primary network is in downlink and the secondary network is in uplink mode, in case of Rayleigh or Rician fading, for $1/2 < f \leq 1$, underlay is asymptotically preferred over TD. When the gains are Nakagami-m distributed, then for $\frac{m}{1+m}< f \leq 1$, simultaneous transmission outperforms TD.

 \item If the primary network is in uplink and the secondary is in downlink mode, in case of Rayleigh or Rician fading, for $0<f \leq 1/2$, underlay is  asymptotically preferred over TD. When the gains have Nakagami-m distribution, then for $0<f\leq \frac{1}{1+m}$, simultaneous transmission outperforms TD.

\item If both networks are in downlink mode, simultaneous transmission asymptotically outperforms TD for any $0< f\leq 1$. In other words, both networks may operate simultaneously in the same band at no loss. This result is achieved when the channel and interference gains have an arbitrary CDF with the exponential tail property.

\end{itemize}

Interestingly, the above enhanced sum-rate results are achievable when uplink scheduling and deactivation of users
is based only on the interference gains to the opposite network. Moreover, in uplink mode, user
 scheduling based on interference to the other network as well as gains to one's own network does
  not improve the results. Therefore by only allowing activation of users that generate interference
   less than a certain threshold, the secondary network can achieve a significant throughput while the primary
    network is protected to the desired protection factor. It can be easily verified that an increase in the variance
    of the Nakagami-m distribution (decrease in $m$) expands the range of $f$ for which better sum-rates are achieved.
    In other words, the richer the scattering environment, the better the asymptotic sum-rates over TD.

The rest of this paper is organized as follows. In the next section, we introduce our system model, notation, and definitions. We also present a key lemma on the maximum of order statistics and a proposition on the sum of lower order statistics. In sections \ref{sec:pusu} to \ref{sec:pdsd}, sum-rates of primary and secondary networks are derived in each of the four uplink-downlink scenarios. We compare the secondary sum-rates in simultaneous transmission to that in TD in section \ref{sec:Compare}. Discussions about the value of the threshold are presented in section \ref{sec:Thresholds} and simulation results on the accuracy of the key lemma and the proposition are provided in section \ref{sec:Simulation}. We conclude this work in section \ref{sec:Conclusion} and discuss some directions for further research. The proof of the key lemma and the proposition are relegated to the appendices.

\section{Preliminaries}
\label{sec:Preleminaries}

\subsection{System Model and Notations}

We consider two collocated point-to-multipoint
networks that share the same frequency band. We assume that the network with priority access to the band (referred to as the primary network) comprises a base station and $n$ users.
The channel power gains between each primary user and the primary base station are denoted by $G^p_i$, $i = 1, \ldots, n$.

Likewise, the second network (referred to as the secondary network) consists of a base station and
$k$ users. Also, we denote by $G^s_j$, $j = 1, \ldots, k$, the channel power gains
between each of the secondary users and the secondary base station.
Throughout the paper, we choose $k = n^{\alpha}$
for some  $\alpha > 0$.

\begin{figure}[t1]
\begin{center}
\epsfig{keepaspectratio = true, width = 3.5 in, figure = 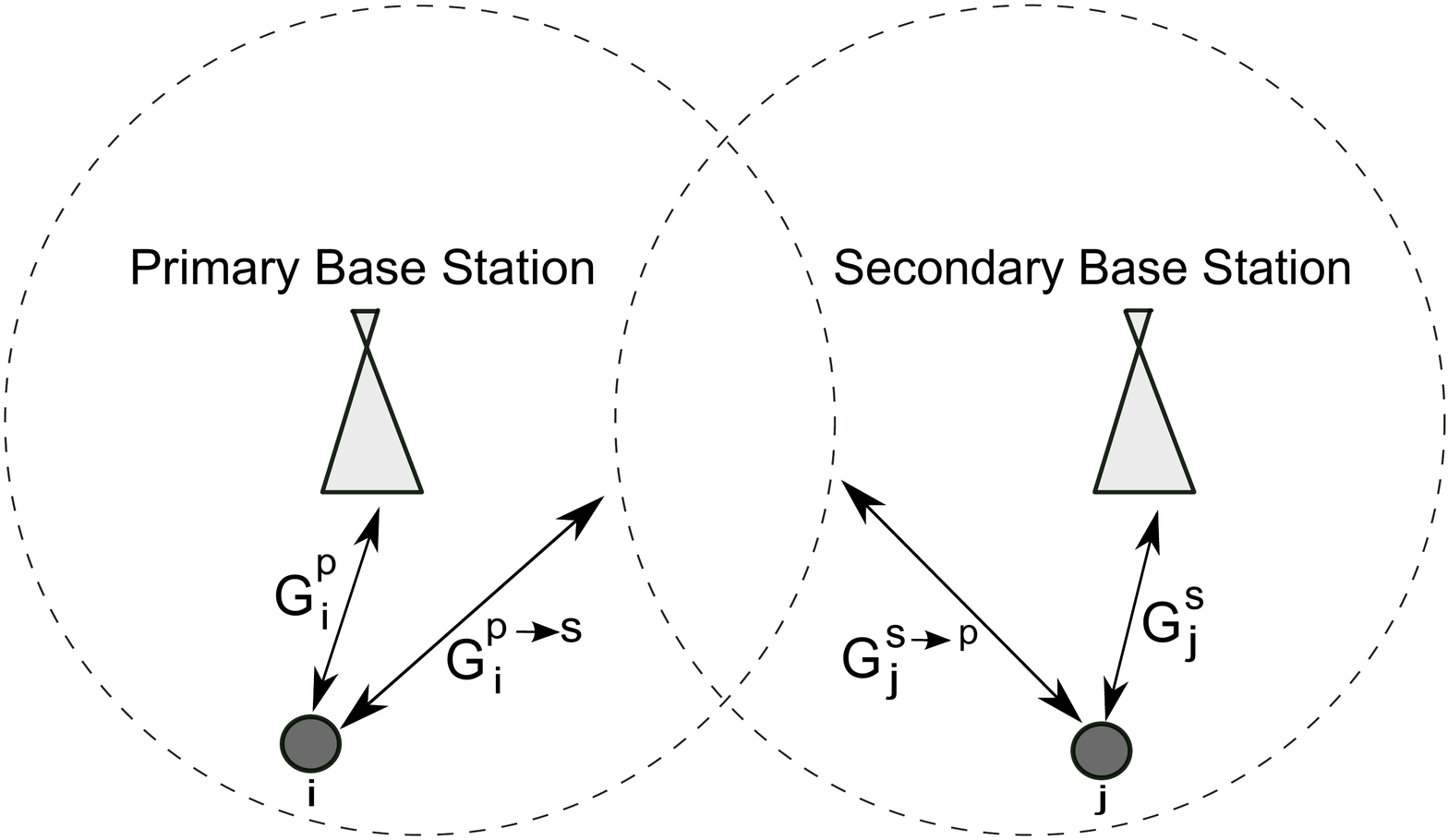}
\end{center}
\caption{Channel power gain between primary's $i$'th (resp. secondary's $j$'th) user and primary (resp. secondary) base station. Interference power gain between primary's $i$'th (resp. secondary's $j$'th) user and secondary (resp. primary) receiver. }
\label{fig:system-model}
\end{figure}

Simultaneous transmission gives rise to $(n+1)(k+1)$ possible
interference gains between the primary and secondary networks. However, at each instant, only one single user or base station will act as a receiver\footnotemark \footnotetext{In uplink, only a base station is receiving,
while in downlink, transmitting to the receiver with
the strongest gain maximizes sum-rate.}. We thus denote by
$G^{s \rightarrow p}_j$ the interference power gain between secondary
user $j$ and the primary receiver. Similarly, the interference power gain
between primary user $i$ and the secondary receiver is denoted by $G^{p \rightarrow s}_i$, with the special
case that $i=0$ (resp. $j=0$) denotes the interference from
the primary (resp. secondary) base station.

We further assume that the interference power gains and channel power gains are i.i.d. with CDF $F_G(g)$.


Fig. \ref{fig:system-model} shows primary and secondary networks, channel power gains, and interference power gains. For simplicity, we illustrate only one user in each network and its corresponding channel and interference power gains.

As stated earlier, each of the primary and secondary networks can operate in
either uplink (i.e., multiple access channel) or
downlink (i.e., broadcast channel) modes, resulting in four
possible scenarios for their simultaneous activity.

In both networks, each transmitter is assumed to employ a complex Gaussian codebook
with power $P$, and each receiver is subject to circularly symmetric additive white
Gaussian noise with power $N$.

Throughout the paper, we denote by $R_\Sigma (n)$ the sum-rate of a single network with $n$ users either in uplink or downlink mode.
When both primary and secondary networks operate in the same band via TD, $R_{\Sigma , \sf{TD}} ^p(n,k)$ denotes the sum-rate of the primary network, while $R_{\Sigma , \sf{TD}} ^s(n,k)$ denotes the sum-rate of the secondary network, where $n$ and $k$ are the number of primary and secondary users, respectively.
 In case of simultaneous transmission, $R_{\Sigma , \sf{Sim}} ^p(n,k)$ and $R_{\Sigma , \sf{Sim}} ^s(n,k)$ represent primary and secondary sum-rates, respectively. 

\subsection{Definitions}
\new{\bf Definition 1:} Let $X_1, X_2, \ldots$ be a sequence
of random variables whose mean
may not converge. We say that the sequence $X_n$ concentrates
as the non-random sequence $a_n>0$ if
\begin{align}
 {\sf Pr}[|X_n - a_n| \geq \epsilon a_n] \rightarrow 0,
\end{align}
as $n \rightarrow \infty$ for all $\epsilon > 0$, i.e. ,
$X_n/a_n$ converges in probability to 1. We denote
this fact by $X_n \dotsim a_n$.

We also abuse notation and write $X_n\dotsim 0$ when $X_n$ converges in probability to zero.
\\ \endproof

{\em Remark 2:} It can be verified that
given positive sequences $a_n$ and $b_n$,
if $X_n \dotsim a_n$ and $\lim_{n\rightarrow \infty} a_n/b_n=1$,
then $X_n \dotsim b_n$. We then write $X_n \dotsim a_n \dotsim b_n$.
Also, if $X_n\dotsim a_n$ and $\lim_{n\rightarrow \infty} a_n=0$, then $X_n\dotsim 0$. We then write $X_n\dotsim a_n\dotsim 0$.\\

 {\em Remark 3:} Let $X_n$ and $Y_n$ be two sequences of random variables where $X_n\dotsim x_n$ and $Y_n\dotsim y_n$ for strictly positive sequences
$x_n$ and $y_n$. Then
\begin{align}
\frac{X_n}{Y_n}\dotsim\frac{x_n}{y_n}\cdot
\end{align}
If the limit, $\lim_{n\rightarrow \infty} x_n/y_n$ exists, then $\frac{X_n}{Y_n}$ converges in probability to that limit, i.e.,
\begin{align}
\plim_{n\rightarrow \infty}\frac{X_n}{Y_n}=\lim_{n\rightarrow \infty}\frac{x_n}{y_n},
\end{align}
where ``$\plim$" refers to limit in probability.

Throughout the paper, the sequences of random variables represent the sum-rates of the networks.
Sum-rates of a single network with $n$ users in uplink or downlink modes are two well-known examples.

{\em Example 4:} (Single Network Uplink) The sum-rate of a Gaussian multiple access channel
with $n$ users is known to be\footnotemark \cite{MAC}
\footnotetext{Throughout the paper, $\log$ is logarithm base-2 while $\ln$ is the natural logarithm.}
\begin{align}
 R_{\Sigma}(n) = \log \left( 1 + \frac{P \sum_{i=1}^n G_i}{N} \right).
\end{align}
Since the channel power gains are \iid with unit mean, then
\begin{align}
R_{\Sigma}(n)\dotsim \log \left( 1 + \frac{P n}{N} \right)\dotsim \log n.
\label{MAC}
\end{align}

{\em Example 5:} (Single Network Downlink) Consider a broadcast channel
with $n$ potential receivers. It is well-known that transmitting
to the receiver with the strongest gain optimizes the
system throughput \cite{broadcast}. Hence, the sum-rate is
\begin{align}
 R_{\Sigma}(n) = \log \left( 1 + \frac{ P\max_{i=1}^n G_i}{N} \right)\cdot
\end{align}

\new{\bf Definition 6:}
Let $X$ be a random variable with probability distribution function (pdf) $f_X{(x)}$. We say $X$ has an exponential tail with parameter $c$ if:
\begin{align}
\lim _{x\rightarrow\infty}\frac {\ln {f_X{(x)}}}{x} = -c ,
\end{align}
where $c$ is a positive real number.\\
\endproof
%

\new{\bf Definition 7:}
Let $X$ be a random variable with CDF $F_X(x)$. We say $X$ has parameters $\lambda>0$ and $\gamma>0$ as $x\rightarrow 0$
if and only if there exist positive numbers $\delta$ and $M$ such that
\begin{align}
\left| F_X(x)-\lambda x^\gamma\right|\leq M \left| x^{\gamma +1} \right|
\end{align}
for every $\left| x\right|<\delta$.
We express this fact by,
\begin{align}
 F_X(x) = \lambda x^\gamma +O(x^{\gamma+1})\cdot
 \label{proposition}
 \end{align}
\endproof

Now, we present some examples of the properties mentioned in definitions 6 and 7 where $X$ is a random variable with ${\sf{E}}[X^2]=2$.

 {\em Example 8:} If $X$ is Rayleigh distributed, then the random variable $G={X^2}/2$ will have an exponential tail with parameter $c=1$ and it also has parameters
$\lambda =1$ and $\gamma =1$ as $g\rightarrow 0$.

{\em Example 9:} If $X$ is Rician distributed with parameter $K$ ($K$-factor) as the ratio of the LOS component's power to the power of the multipath fading component, then $G={X^2}/2$ will have an exponential tail with parameter $c=K+1$ and parameters
$\lambda =(1+K) e^{-K}$ and $\gamma =1$ as $g\rightarrow 0$.

{\em Example 10:} If $X$ is Nakagami distributed, with parameter $m$ (Nakagami-m distributed), then $G={X^2}/2$ will have an exponential tail with $c=m$ and parameters
$\lambda = \frac{m^{m-1}}{\Gamma(m)}$ and $\gamma =m$ as $g\rightarrow 0$.

Proofs of the three examples are provided in Appendix A.

\subsection{Proposition and Lemma}

Given a set of random variables
$X_1, X_2, \ldots, X_n$, we denote its order statistics by $X_{1:n}, \ldots, X_{n:n}$, i.e., $X_{1:n} \leq X_{2:n} \leq \cdots \leq X_{n:n}$.

In the following, we have a lemma on the asymptotic behavior of the maximum of $n$ \iid random variables and a proposition on the asymptotic sum of lower order statistics, both playing key roles in the sequel.\\

\new{\bf Lemma 11:}
Let $X_1,X_2 ,\ldots, X_n $ be \iid random variables with pdf $f_{X}{(x)}$ and an exponential tail with parameter $c>0$.
Define $X_{n:n}=\max_{i=1}^n X_i$. Then,
\begin{align}
X_{n:n}\dotsim \frac {\ln n}{c} \cdot
\end{align}
 \endproof

\new{\bf Proposition 12:}
Let $X_1, \ldots, X_n$ be \iid random variables with unit mean, CDF $F_X(x)$ and parameters $\lambda$ and $\gamma$ as $x\rightarrow 0$.
Furthermore, let $f : \naturals \rightarrow \naturals$ be such that $f(n) \rightarrow \infty$
and $f(n)/n \rightarrow 0$ as $n \rightarrow \infty$.
Then the sum of the $f(n)$ lowest order statistics
\begin{align}
 S_n = \sum_{i=1}^{f(n)} X_{i:n}
\end{align}
concentrates as
\begin{align}
S_n \dotsim \frac{n{(\frac{f(n)}{n})}^{1+\frac{1}{\gamma}}}{\lambda^{\frac{1}{\gamma}}(1+\frac{1}{\gamma})} \cdot
\end{align}
\endproof

Proofs of Lemma 11 and Proposition 12 are provided in Appendix B.

Now recall Example 5 and assume the channel gains have an exponential tail with parameter $c>0$, then by Lemma 11, the network's sum-rate in downlink mode concentrates as
 \begin{align}
 R_\Sigma(n)=\log \left( 1+\frac{P \cdot \frac{\ln n}{c}}{N}\right)\dotsim \log{\log n}\cdot
 \end{align}
\subsection{Baseline Comparison}

As a baseline, we consider channel sharing via TD and we define the primary protection factor $f$ as the limit
\begin{align}
f= \plim_{n\rightarrow\infty } \frac{R_{\Sigma , \sf{TD}}^p (n,k)}{R_{\Sigma}(n)},
\end{align}
i.e., it represents the fraction of time $0<f\leq1$ that the primary network uses the channel.
If we take $k = n^\alpha$ ($\alpha>0$) and let the channel and interference power gains have an exponential tail with parameter $c>0$, then using Lemma 11, we obtain asymptotic sum-rates listed in table \ref{t:baseline} for the channel sharing via TD.
We refer to the limit
\begin{align}
\plim_{n\rightarrow\infty} \frac{R_{\Sigma , \sf{TD}}^s(n,n^\alpha)}{R_{\Sigma}(n^\alpha)}
\end{align}
as the secondary throughput factor in TD which is easily verified to be equal to $1-f$.

Unsurprisingly, if the primary network is not to be impacted at all
(i.e., $f=1$), then the secondary system is silent.

\begin{table}
\center
 \caption{Asymptotic primary and secondary sum-rates in TD.}
 \label{t:baseline}
\begin{tabular}{cccc}
\toprule
Primary & Secondary & $R_{\Sigma , \sf{TD}}^p(n,n^\alpha)$ & $R_{\Sigma , \sf{TD}}^s(n, n^\alpha)$ \\
\hline
up   & up   & $f \log n$      & $\alpha(1-f) \log n$ \\
up   & down & $f \log n$      & $(1-f) \log \log n$ \\
down & up   & $f \log \log n$ & $\alpha(1-f) \log n$ \\
down & down & $f \log \log n$ & $(1-f) \log \log n$ \\
\bottomrule
\end{tabular}
\end{table}

\subsection{CSI and Co-existence}

In downlink mode, CSI refers to the knowledge of a network's own channel power gains which is available to the network's base station. When the network operates in downlink mode with such CSI, the highest sum-rate is obtained by transmitting to the user with the greatest channel power gain (the strongest user). Knowledge of the interfering channel gains between the base station and the opposite network's receiver has no influence on which receiver is scheduled.

In uplink mode, CSI refers to the knowledge of a network's own channel power gains as well as the knowledge of the interference power gains between the users in the network and the other network's receiver. This information is available to the network's scheduler, which applies the CSI to activate or deactivate users of the network so that the network obtains the highest sum-rate and also the interference imposed on the other network is minimized.

For uplink transmission, three different levels of co-existence are considered based on utilization of the CSI.

{\em Pure Interference:} At this simplest level of co-existence, the primary and secondary networks are considered as
two independent entities. None of the networks employ the CSI to schedule their users. Therefore, all the primary and secondary users are active and interfere with each other.\\

{\em Asymmetric co-existence:} At the asymmetric co-existence level, only the secondary network schedules its users.
 In other words, only the secondary network's scheduler selectively activates or deactivates secondary users. The primary network however, operates as usual.\\

{\em Symmetric co-existence:} The third level is referred to as symmetric co-existence level.
In this case, the primary network also applies its CSI and selectively schedules its users.
\subsection{Scheduling Strategies}

In uplink transmission, the primary and secondary users can be activated or deactivated based on two different scheduling strategies which are referred to as joint optimization, and least interference strategies. These strategies are performed by the scheduler in each network.

 Using the joint optimization strategy, the scheduler of a network can optimally deactivate the network's users based on the network's own channel power gains and the interference power gains between its users and the opposite network's receiver (CSI). The joint optimization scheduling can be performed in either of the networks such that the highest secondary throughput factor is achieved while the primary network is protected.

However, when using the least interference scheduling method, activation or deactivation of users is based only on their interference power gains to the opposite network and not on the networks' own channel power gains. In other words, the scheduler only considers the interference power gains rather than the network's own channel power gains.

For the pure interference as well as each scheduling strategy in the symmetric or asymmetric co-existence levels, we are interested in maximizing the secondary throughput factor defined as
\begin{align}
\plim_{n\rightarrow \infty}\frac{R_{\Sigma , \sf{Sim}}^s (n,n^ \alpha)}{R_{\Sigma}(n^\alpha)},
\label{secfact}
\end{align}
subject to the constraint,
\begin{align}
\label{pusu2}
\plim_{n\rightarrow \infty}\frac{R_{\Sigma , \sf{Sim}}^p (n,n^ \alpha)}{R_{\Sigma}(n)} \geq f
\end{align}
for some primary protection factor $0<f \leq 1$.

In what follows, we examine each of the uplink-downlink scenarios and find the asymptotic sum-rates at each co-existence level.
\section{Primary and Secondary Uplink}
\label{sec:pusu}
In this section, we derive the asymptotic primary and secondary sum-rates under simultaneous transmission when both networks operate in uplink mode. We examine the three different co-existence levels, and, at each level, we find the asymptotic result for the maximum secondary throughput factor provided that the primary protection factor be at least $f$.

\subsection{Pure Interference}
We first consider the pure interference case. Having $n$ and $n^\alpha$ ($\alpha>0$) active primary and secondary users respectively, the sum-rates of the primary and secondary networks are,
\begin{align}
 R_{\Sigma , \sf{Sim}}^p(n, n^\alpha) = \log \left( 1 + \frac{P \sum_{i=1}^n G_i^p}{N + P \sum_{j=1}^{n^\alpha} G^{s\rightarrow p}_j} \right),
\end{align}
and
\begin{align}
 R_{\Sigma , \sf{Sim}}^s(n, n^\alpha) = \log \left( 1 + \frac{P \sum_{j=1}^{n^\alpha} G_j^s}{N + P \sum_{i=1}^n G^{p\rightarrow s}_i} \right),
\end{align}
respectively. For primary and secondary uplink scenario, $G^{s\rightarrow p}_j$ is the interference power gain from secondary user $j$ to the primary base station and $G^{p\rightarrow s}_i$ is the interference power gain from primary user $i$ to the secondary base station.

For any $0<f\leq1$, \eqref{pusu2} should be satisfied in order to protect the primary network.
Since
\begin{align}
 R_{\Sigma , \sf{Sim}}^p (n,n^\alpha)\dotsim \log \left( 1 + \frac{P n}{N + P {n^\alpha}} \right) \dotsim \log {n^{1-\alpha}},
\end{align}
for $1-\alpha >0$, and
\begin{align}
 R_{\Sigma}(n) \dotsim \log \left( 1 + \frac{P n}{N} \right) \dotsim \log {n},
\end{align}
the primary protection constraint, \eqref{pusu2}, cannot be satisfied for any value $\alpha>0$ when $f=1$. For $0<f<1$,
the primary network is protected when $1-\alpha \geq f$ which results in
\begin{align}
R_{\Sigma , \sf{Sim}}^s(n, n^\alpha) \dotsim \log \left( 1 + \frac{P {n^\alpha}}{N + P {n}} \right) \dotsim 0\cdot \end{align}
Therefore,
\begin{align}
\max _{0< \alpha \leq 1-f}\left[\plim_{n\rightarrow \infty}\frac{R_{\Sigma, \sf{Sim}}^s (n,n^\alpha)}{R_{\Sigma}(n^\alpha)}\right]= 0 .
\end{align}
for any $0<f<1$.

This negative result is not surprising, because the secondary network has no
method to limit its interference to the primary network, except by decreasing the number of active users.

Thus, at the pure interference level, no better secondary sum-rate can be achieved over TD.
\subsection{Co-existence}
Using CSI, the secondary and primary networks can sufficiently mitigate the interference to the opposite
network by deactivating some of their users. Therefore, a positive secondary throughput factor may be achieved while the primary network is protected.

 First, we examine the joint optimization strategy. In this case, if
the primary network activates $n^\beta$, $0 \leq \beta \leq 1$, users out of $n$ and the secondary network activates $n^{\bar \alpha} $, users out of $n^{\alpha}$, $0 \leq \bar{\alpha} \leq \alpha $, then the sum-rates of the primary and secondary networks can be written as follows,
\begin{align}
R_{\Sigma , \sf{Sim}}^p (n,n^ \alpha) = \log \left( 1 + \frac{P\sum_{\ell=1}^{n^\beta} G^p_{i_\ell}} {N + P\sum_{\bar \ell=1}^{n^{\bar \alpha}} G^{s\rightarrow p}_{j_{\bar \ell}}} \right),
\end{align}
and
\begin{align}
R_{\Sigma , \sf{Sim}}^s (n,n^ \alpha) = \log \left( 1 + \frac{P\sum_{\bar \ell=1}^{n^{\bar \alpha}} G^s_{j_{\bar \ell}}} {N + P\sum_{\ell=1}^{n^\beta} G^{p\rightarrow s}_{i_{\ell}}} \right)\cdot
\end{align}

The joint optimization is over $\bar \alpha$, $\beta$, and the set of vectors $(j_1, j_2,\ldots, j_{n^{\bar \alpha} })$ and $(i_1, i_2,\ldots, i_{n^\beta})$, such that the primary is protected and the secondary throughput factor is maximized.

  On the other hand, following the least interference strategy, the primary network activates $n^\beta$ users, $0 \leq \beta \leq 1$, out of $n$ that result in the least interference power gains to the secondary receiver and similarly, the secondary network activates  $n^{\bar{\alpha}}$ users out of $n^{\alpha}$, $0 \leq \bar{\alpha} \leq \alpha $, which generate the least interference power gains to the primary receiver.
 Using this strategy, the sum-rates of primary and secondary networks can be written as
\begin{align}
R_{\Sigma , \sf{Sim}}^p (n,n^ \alpha) = \log \left( 1 + \frac{P\sum_{\ell=1}^{n^\beta} G^p_{i_\ell}} {N + P\sum_{j=1}^{n^{\bar{\alpha}}} G^{s\rightarrow p}_{j:{n^\alpha}}} \right),
\end{align}
and
\begin{align}
R_{\Sigma , \sf{Sim}}^s (n,n^ \alpha) = \log \left( 1 + \frac{P\sum_{\ell=1}^{n^{\bar{\alpha}}} G^s_{j_\ell}}
                               {N + P \sum_{i=1}^{n^{\beta}} G^{p\rightarrow s}_{i:n}} \right).
\end{align}

\new{\bf Theorem 13:}
Let interference and channel power gains be i.i.d. random variables with unit mean and an exponential tail with parameter $c>0$ and parameters $\lambda>0$ and $\gamma>0$. Given the protection of the primary network and using the joint optimization strategy, an upper bound on the maximum achievable secondary throughput factor at the asymmetric co-existence level is
 $\left[\frac{\alpha -1 -f \gamma }{\alpha (1+\gamma)}\right]^+$ for every $0<f \leq 1$ and $\alpha>0$. In case of symmetric co-existence, for $0<f\leq \frac{1}{1+\gamma}$ we have the upper bound
\begin{align}
\left\{
\begin{array}{cl}
\left[\frac{\gamma}{\alpha (\gamma +1)^2} -(\frac{\gamma}{\gamma +1})\frac{f}{\alpha} +\frac{1}{1+\gamma}\right]^+, & \alpha \geq \frac{1}{1+\gamma}-f\\
1, &0<\alpha < \frac{1}{1+\gamma}-f,
\end{array}
\right.
\end{align}
and for $\frac{1}{1+\gamma}<f\leq 1$, we have the upper bound
\begin{align}
 \left[\frac{1}{\alpha \gamma}- (1+\frac{1}{\gamma})\frac{f}{\alpha} + \frac{1}{1+\gamma}\right]^+,
\end{align}
for $\alpha>0$, where $x^+\eqdef\max(0,x)$. Furthermore, these bounds are achievable by using the least interference scheduling strategy.
\endproof

 \new{\bf Proof of Theorem 13:} In the first part of the proof, we show that the above result is achievable when scheduling users is based only on the least interference power gains.

  Letting $F_G(g)$ be the CDF of the channel and interference power gains, then $F_G(g)=\lambda g^\gamma +O(g^{\gamma+1})$ as $g\rightarrow 0$. Using Proposition 12, for any $\alpha>0$ we have
\begin{align}
  \sum_{j=1}^{n^{\bar{\alpha}}} G_{j:n^\alpha}^{s \rightarrow p}\dotsim \frac{n^{\bar{\alpha} (1+\frac{1}{\gamma})-\frac{\alpha}{\gamma}}}{\lambda ^{\frac{1}{\gamma}}(1+\frac{1}{\gamma})},
  \end{align}
  and
  \begin{align}
  \sum_{i=1}^{n^{\beta}} G_{i:n}^{p \rightarrow s}\dotsim \frac{n^{\beta (1+\frac{1}{\gamma})-\frac{1}{\gamma}}}{\lambda ^{\frac{1}{\gamma}}(1+\frac{1}{\gamma})}\cdot
  \end{align}
Thus, sum-rate of the primary network concentrates as
   \begin{align}
  R_{\Sigma , \sf{Sim}}^p (n,n^\alpha) &\dotsim \log \left( 1 + \frac{P {n^\beta} }{N + P \cdot \frac{n^{{\bar {\alpha}} (1+\frac{1}{\gamma})-\frac{\alpha}{\gamma}}}{\lambda ^{\frac{1}{\gamma}}(1+\frac{1}{\gamma})}} \right)\label{upper3}\\
  &\dotsim  \left(\beta - \left[( 1+ \frac{1}{\gamma}) \bar{\alpha} - \frac{\alpha}{\gamma}\right]^+ \right)^+ \log n .
\end{align}
Likewise, the sum-rate of the secondary network concentrates as
\begin{align}
  R_{\Sigma , \sf{Sim}}^s (n,n^\alpha) &\dotsim \log \left( 1 + \frac{P {n^{\bar \alpha}} }{N + P \cdot \frac{n^{{\beta} (1+\frac{1}{\gamma})-\frac{1}{\gamma}}}{\lambda ^{\frac{1}{\gamma}}(1+\frac{1}{\gamma})}} \right)\label{upper4}\\ &\dotsim  \left({\bar \alpha} - \left[( 1+ \frac{1}{\gamma}) {\beta} - \frac{1}{\gamma}\right]^+ \right)^+ \log n .
\end{align}
%

We now consider two co-existence levels:
\begin{itemize}
\item {{\em Case $\beta = 1$ (Asymmetric Co-existence)}:
In this case, we have to maximize $\left[{{\bar \alpha} -1}\right]^+$
subject to the primary protection constraint
\begin{align}
1 - \left[( 1+ \frac{1}{\gamma}) \bar{\alpha} - \frac{\alpha}{\gamma}\right]^+ \geq f ,
\label{constraint1}
\end{align}
for every $0<f\leq 1$, as well as $0 \leq \bar{\alpha} \leq \alpha$.

Solving the optimization problem when $\alpha>1$, we find $\bar{\alpha} = \frac{\gamma-f \gamma+\alpha}{1+\gamma}$ as an optimal value. In the case that $\alpha\leq 1$, the secondary sum-rate trivially concentrates as $0$.
 Therefore, the maximum secondary throughput factor is
\begin{align}
\max _{\bar \alpha}\left[\plim_{n\rightarrow \infty}\frac{R_{\Sigma , \sf{Sim}}^s (n,n^ \alpha)}{R_{\Sigma}(n^\alpha)}\right]
= \left[\frac{\alpha -1 -f \gamma }{\alpha (1+\gamma)}\right]^+
\end{align}
for any $0<f\leq 1$ and $\alpha>0$.
}
\item {{\em Case $0 \leq \beta < 1$ (Symmetric Co-existence)}:
In this case, we must maximize $\left[{\bar \alpha} - \left[( 1+ \frac{1}{\gamma}) {\beta} - \frac{1}{\gamma}\right]^+\right]^+$
subject to the primary protection constraint
\begin{align}
\beta - \left[( 1+ \frac{1}{\gamma}) \bar{\alpha} - \frac{\alpha}{\gamma}\right]^+ \geq f ,
\end{align}
 for every $0<f\leq 1$, as well as $0 \leq \bar{\alpha} \leq \alpha$.

For $0<f\leq \frac{1}{1+\gamma}$, the optimal solution is to select
\begin{align}
\left\{
\begin{array}{ll}
\beta =\frac{1}{1+\gamma}, {\bar \alpha}= \frac{\gamma(\beta -f) +\alpha}{1+\gamma}, & \alpha \geq \frac{1}{1+\gamma}-f \\
\beta =\frac{1}{1+\gamma}, {\bar \alpha}=\alpha, & 0<\alpha < \frac{1}{1+\gamma}-f\cdot
\end{array}
\right.
\end{align}

For $ \frac{1}{1+\gamma}< f \leq 1$, the optimal solution is to select $\beta =f$ and ${\bar \alpha}=\frac{\alpha}{1+\gamma}$
when $\alpha> \frac{(1+\gamma)^2}{\gamma}f-\frac{1+\gamma}{\gamma}$. In the case that $\alpha\leq \frac{(1+\gamma)^2}{\gamma}f-\frac{1+\gamma}{\gamma}$ the secondary sum-rate trivially concentrates as $0$.

Thus, for $0<f\leq \frac{1}{1+\gamma}$ the maximum secondary throughput factor is
\begin{align}
\left\{
\begin{array}{cl}
\left[\frac{\gamma}{\alpha (\gamma +1)^2} -(\frac{\gamma}{\gamma +1})\frac{f}{\alpha} +\frac{1}{1+\gamma}\right]^+, & \alpha \geq \frac{1}{1+\gamma}-f\\
1, &\alpha < \frac{1}{1+\gamma}-f,
\end{array}
\right.
\end{align}

and for $\frac{1}{1+\gamma}<f\leq 1$, the maximum secondary throughput factor is
\begin{align}
 \left[\frac{1}{\alpha \gamma}- (1+\frac{1}{\gamma})\frac{f}{\alpha} + \frac{1}{1+\gamma}\right]^+,
\end{align}
when $\alpha>0$.
}
\end{itemize}

 The final step of the proof is to show that using joint optimization, we achieve the same results. Scheduling users based on joint optimization, one can find upper bounds for primary and secondary sum-rates as follows
 \begin{align}
 R_{\Sigma , \sf{Sim}}^p (n,n^ \alpha) \leq R_{\Sigma , \sf{UB}, \sf{Sim}}^p (n,n^ \alpha),
\end{align}
and
\begin{align}
 R_{\Sigma , \sf{Sim}}^s (n,n^ \alpha) \leq R_{\Sigma , \sf{UB}, \sf{Sim}}^s (n,n^ \alpha),
\end{align}
 where
  \begin{align}
 R_{\Sigma , \sf{UB}, \sf{Sim}}^p (n,n^ \alpha) = \log \left( 1 + \frac{P n^\beta  G^p_{n:n}} {N + P \sum_{j=1}^{n^{\bar \alpha}} G^{s\rightarrow p}_{j:{n^\alpha}} }\right),
 \end{align}
 and
  \begin{align}
R_{\Sigma , \sf{UB}, \sf{Sim}}^s (n,n^ \alpha) = \log \left( 1 + \frac{P {n^{\bar \alpha}} G^s_{{n^\alpha}:{n^\alpha}}} {N + P \sum_{i=1}^{n^\beta} G^{p\rightarrow s}_{i:n}} \right)\cdot
\end{align}

These upper bounds are obtained by assuming all activated users have the maximum channel power gains to their own networks and the least interference power gains to the opposite network.

Since the channel power gains have exponential tail with parameter $c>0$, we have
\begin{align}
R_{\Sigma , \sf{UB}, \sf{Sim}}^p (n,n^ \alpha)&\dotsim \log \left( 1 + \frac{P  n^\beta \cdot \frac{\ln n}{c}} {N + P \cdot \frac{n^{{\bar \alpha}(1+\frac{1}{\gamma})} n^{-\frac{\alpha}{\gamma}}}{\lambda^{\frac{1}{\gamma}}(1+\frac{1}{\gamma})} }\right)\nonumber \\& \dotsim \log \left( 1 + \frac{P {n^\beta} }{N + P \cdot \frac{n^{{\bar {\alpha}} (1+\frac{1}{\gamma})-\frac{\alpha}{\gamma}}}{\lambda ^{\frac{1}{\gamma}}(1+\frac{1}{\gamma})}} \right),
\label{upper1}
\end{align}
and
\begin{align}
R_{\Sigma , \sf{UB}, \sf{Sim}}^s (n,n^ \alpha)&\dotsim \log \left( 1 + \frac{P {n^{\bar \alpha}}\cdot \frac{\ln {n^\alpha}}{c}} {N + P \cdot \frac{n^{{\beta}(1+\frac{1}{\gamma})} n^{-\frac{1}{\gamma}}}{\lambda^{\frac{1}{\gamma}}(1+\frac{1}{\gamma})} }\right)\nonumber \\ & \dotsim \log \left( 1 + \frac{P {n^{\bar \alpha}} }{N + P \cdot \frac{n^{{\beta} (1+\frac{1}{\gamma})-\frac{1}{\gamma}}}{\lambda ^{\frac{1}{\gamma}}(1+\frac{1}{\gamma})}} \right).
\label{upper2}
\end{align}
But \eqref{upper1} and \eqref{upper2} are identical to \eqref{upper3} and \eqref{upper4} respectively, thus by finding
\begin{align}
\max_{{\bar \alpha},\beta}\left[\plim_{n\rightarrow \infty}\frac{R_{\Sigma , \sf{UB}, \sf{Sim}}^s (n,n^ \alpha)}{R_{\Sigma}(n^\alpha)}\right],
\label{joint:pusu2}
\end{align}
subject to
\begin{align}
\plim_{n\rightarrow \infty}\frac{R_{\Sigma , \sf{UB}, \sf{Sim}}^p (n,n^ \alpha)}{R_{\Sigma}(n)} \geq f,
\label{joint:pusu1}
\end{align}
for $0<f \leq 1$, we obtain the upper bound for the maximum secondary throughput factor to be equal to the maximum secondary throughput factor when user scheduling is based on the least interference strategy. Therefore, by using the joint optimization strategy, the secondary network can not achieve a better sum-rate than that achieved by using the least interference strategy.\\
  \endproof

  The results obtained in this section indicate that, at asymmetric and symmetric co-existence levels, it
is possible for the secondary network to provide positive asymptotic sum-rate while the primary is still
protected by factor $f$, whereas at the pure interference level, the secondary
is asymptotically unable to deliver a positive sum-rate.

\section{Primary Downlink and Secondary Uplink}
\label{sec:pdsu}
As the primary network is in downlink mode, there is no symmetric co-existence for this scenario to be considered.
\subsection{Pure Interference}
In this case, all $n^\alpha$ ($\alpha>0$) secondary users are active and the primary base station transmits to the primary user with the highest channel power gain. Therefore, the sum-rates of primary and secondary networks are
\begin{align}
  R_{\Sigma , \sf{Sim}}^p (n,n^ \alpha) = \log \left( 1 + \frac{P\max_{i=1}^{n} G_i^p}{N + P \sum_{j=1}^{n^\alpha} G_j^{s \rightarrow p}} \right),
\end{align}
and
\begin{align}
R_{\Sigma , \sf{Sim}}^s (n,n^ \alpha) = \log \left( 1 + \frac{P\sum_{j=1}^{n^\alpha} G_j^s}{N + P G_0^{p \rightarrow s}} \right),
\end{align}
respectively.

Let the channel power gains have an exponential tail with parameter $c>0$. Then, by Lemma 11
\begin{align}
\max_{i=1}^{n} G_i^p \dotsim \frac{\ln {n}}{c} \cdot
\end{align}
Thus,
\begin{align}
  R_{\Sigma , \sf{Sim}}^p (n,n^ \alpha) \dotsim \log \left( 1 + \frac{P \cdot \frac{\ln {n}}{c}}{N + P n^\alpha} \right)\dotsim 0,
\end{align}
for any $\alpha>0$.

Due to the excessive interference from the secondary network to the primary receiver, the primary protection
constraint, \eqref{pusu2}, can not be satisfied for any value of $\alpha>0$ and therefore the networks can not coexist at this level.
%
\subsection{Asymmetric co-existence}
 Using the joint optimization scheduling strategy in the secondary network with $n^{\bar \alpha}$ ($0\leq {\bar \alpha}\leq \alpha$) secondary active users and letting the primary base station transmit to the strongest primary user, sum-rates of the primary and secondary networks can be written as
\begin{align}
R_{\Sigma , \sf{Sim}}^p (n,n^\alpha)= \log \left( 1 + \frac{P\max_{i=1}^{n} G^p_{i}} {N + P\sum_{\bar \ell=1}^{n^{\bar \alpha}} G^{s\rightarrow p}_{j_{\bar \ell}}} \right),
\end{align}
and
\begin{align}
R_{\Sigma , \sf{Sim}}^s (n,n^\alpha) = \log \left( 1 + \frac{P\sum_{\bar \ell=1}^{n^{\bar \alpha}} G^s_{j_{\bar \ell}}} {N + P G_0^{p\rightarrow s}} \right),
\end{align}
respectively.
 To solve the above joint optimization problem, we should find ${\bar \alpha}$ and the vector $(j_1, j_2,\ldots, j_{n^{\bar \alpha}})$ such that the primary network is protected and the secondary throughput factor is maximized.

Using the least interference strategy and letting the secondary network activate only $n^{\bar \alpha}$ users that generate the least interference power gains to the primary receiver, sum-rates of the primary and secondary networks are as follows
\begin{align}
  R_{\Sigma , \sf{Sim}}^p (n,n^\alpha) = \log \left( 1 + \frac{P\max_{i=1}^{n} G_i^p}{N + P \sum_{j=1}^{n^{\bar{\alpha}}} G_{j:n^\alpha}^{s \rightarrow p}} \right),
\end{align}
and
\begin{align}
R_{\Sigma , \sf{Sim}}^s (n,n^\alpha) = \log \left( 1 + \frac{P\sum_{\bar \ell=1}^{n^{\bar \alpha}} G^s_{j_{\bar \ell}}} {N + P G_0^{p\rightarrow s}} \right).
\end{align}

Given a scheduling strategy, we are interested in finding the maximum secondary throughput factor while ensuring the primary is protected
by a protection factor at least equal to $f$.\\

\new{\bf Theorem 14:} Let the interference and channel power gains be i.i.d. random variables with unit mean and an exponential tail with parameter $c>0$ and parameters $\lambda>0$ and $\gamma>0$. At the asymmetric co-existence level, provided the protection of the primary network and using the joint optimization scheduling strategy, an upper bound on the maximum secondary throughput factor is $\frac{1}{1+\gamma}$ for any $0<f \leq 1$ and $\alpha>0$.
Furthermore, this bound can be achieved by using the least interference scheduling strategy.\\
\endproof

\new{\bf Proof of Theorem 14:}
First, we examine sum-rate results when user scheduling is based only on the least interference power gains. Since the channel and interference
power gains have an exponential tail with parameter $c>0$ and parameters $\lambda>0$ and $\gamma$, by Lemma 11 and Proposition 12, sum-rates of the primary and secondary networks
 concentrate as
   \begin{align}
  R_{\Sigma , \sf{Sim}}^p (n,n^\alpha) &\dotsim \log \left( 1 + \frac{P \cdot \frac{\ln {n}}{c} }{N + P \cdot \frac{n^{{\bar {\alpha}} (1+\frac{1}{\gamma})-\frac{\alpha}{\gamma}}}{\lambda ^{\frac{1}{\gamma}}(1+\frac{1}{\gamma})}} \label{upper7}\right).
\end{align}
and
\begin{align}
R_{\Sigma , \sf{Sim}}^s (n,n^\alpha) \dotsim  \log \left( 1 + \frac{P {n^{\bar \alpha}}}{N + P G_0^{p\rightarrow s}} \right)\dotsim  {{\bar\alpha} \log n}.\label{upper8}
\end{align}
respectively.

To obtain the highest secondary throughput factor, we must maximize ${\bar{\alpha}}$ subject to the primary protection constraint \eqref{pusu2}.

In \eqref{upper7}, when $\bar{\alpha} >\alpha/(1+\gamma)$, the sum-rate of the primary network concentrates as $0$. On the other hand, in the case that $\bar{\alpha} \leq\alpha/(1+\gamma)$, \eqref{pusu2} will be satisfied for every $0< f\leq 1$. Thus, $\bar{\alpha} =\alpha/(1+\gamma)$ is the optimal value and the maximum secondary throughput factor is then
\begin{align}
 \max_{\bar \alpha} \left[\plim_{n\rightarrow \infty}\frac{R_{\Sigma , \sf{Sim}}^s (n,n^\alpha)}{R_{\Sigma}(n^\alpha)}\right]=\frac{1}{1+\gamma},
 \end{align}
 for any $0<f \leq 1$ and $\alpha>0$.
Hence, the first part of Theorem 14 is proven.

In what follows, we show that applying the joint optimization strategy, the sum-rate results will be the same in the best case.

If users are scheduled based on the joint optimization strategy, we will have
  \begin{align}
 R_{\Sigma , \sf{UB}, \sf{Sim}}^p (n,n^ \alpha) = \log \left( 1 + \frac{P\max_{i=1}^{n} G^p_{i}} {N + P\sum_{j=1}^{n^{\bar \alpha}} G^{s\rightarrow p}_{j:n^\alpha}} \right),
 \end{align}
 and
  \begin{align}
R_{\Sigma , \sf{UB}, \sf{Sim}}^s (n,n^ \alpha) = \log \left( 1 + \frac{P  n^{\bar \alpha}  G^s_{n^\alpha:n^\alpha}} {N + P  G_0^{p\rightarrow s}} \right),
\end{align}
as the upper bounds for the primary and secondary sum-rates respectively. These upper bounds are derived by assuming that the secondary active users have the highest channel power gains to their own base station and they also have the least interference power gains to the primary receiver.

Using Lemma 11 and Proposition 12, we have
\begin{align}
R_{\Sigma , \sf{UB}, \sf{Sim}}^p (n,n^ \alpha)&\dotsim \log \left( 1 + \frac{P \cdot \frac{\ln n}{c}} {N + P \cdot \frac{n^{{\bar \alpha}(1+\frac{1}{\gamma})} n^{-\frac{\alpha}{\gamma}}}{\lambda^{\frac{1}{\gamma}}(1+\frac{1}{\gamma})} }\right)\label{upper5},
\end{align}
and
\begin{align}
R_{\Sigma , \sf{UB}, \sf{Sim}}^s (n,n^ \alpha)&\dotsim \log \left( 1 + \frac{P {n^{\bar \alpha}} \cdot \frac{\ln {n^\alpha}}{c}} {N + P G_0^{p \rightarrow s}} \right)\nonumber \\ & \dotsim {\bar \alpha}\log n \label{upper6}\cdot
\end{align}

 Since \eqref{upper5} and \eqref{upper6} are identical to \eqref{upper7} and \eqref{upper8} respectively, by maximizing \eqref{joint:pusu2} over $\bar {\alpha}$ subject to \eqref{joint:pusu1}, we find that
 the maximum secondary throughput factor is upper bounded by that achieved in the least interference strategy.
\endproof

\section{Primary Uplink and Secondary Downlink}
\label{sec:pusd}
As the secondary network is in downlink mode, there is no asymmetric co-existence for this scenario to be considered.
\subsection{Pure Interference}
In this case, following the same line of development as in the first two scenarios,
we find
\begin{align}
R_{\Sigma , \sf{Sim}}^p (n,n^\alpha) \dotsim \log \left( 1 + \frac{P n}{N + P G_0^{s\rightarrow p}} \right) \dotsim \log {n},
\end{align}
and
\begin{align}
 R_{\Sigma}(n) \dotsim \log \left( 1 + \frac{P n}{N} \right) \dotsim \log {n}.
\end{align}
Thus, \eqref{pusu2} is satisfied for any $0<f \leq 1$.

By Lemma 11,
\begin{align}
R_{\Sigma , \sf{Sim}}^s (n,n^\alpha) \dotsim \log \left( 1 + \frac{P \cdot \frac{\ln {n^\alpha}}{c}}{N + P n} \right)\dotsim 0,
\end{align}
for any $\alpha>0$ and thus
\begin{align}
\max_{\alpha>0} \left[\plim_{n\rightarrow \infty}\frac{R_{\Sigma , \sf{Sim}}^s (n,n^\alpha)}{R_{\Sigma}(n^\alpha)}\right]= 0,
\end{align}
for any $0<f \leq 1$.
Hence, the maximum secondary throughput factor is equal to zero in the pure interference case.

\subsection{Symmetric co-existence}
%
%
\new{\bf Theorem 15:}
Let the interference and channel power gains be i.i.d. random variables with unit mean and an exponential tail
with parameter $c>0$ and parameters $\lambda>0$ and $\gamma>0$. For any $\alpha>0$, given the protection of the primary network and
 using the joint optimization strategy, an upper bound on the maximum secondary throughput factor is $1$ for any
  $ 0<f \leq \frac{1}{1+\gamma}$ and $0$ for any $  \frac{1}{1+\gamma}<f\leq 1$. Furthermore, this result can be
  achieved by using the least interference scheduling strategy.\\
\endproof

 \new{\bf Proof of Theorem 15:} When scheduling users is based on the least interferences, the primary sum-rate concentrates as
\begin{align}
R_{\Sigma , \sf{Sim}}^p (n,n^\alpha) \dotsim  \log \left( 1 + \frac{P {n^\beta}}{N + P G_0^{s\rightarrow p}} \right)\dotsim \beta \log n .
\end{align}
for $0<\beta\leq 1$.

Using Lemma 11 and Proposition 12, we find that
   \begin{align}
  R_{\Sigma , \sf{Sim}}^s (n,n^\alpha) &\dotsim \log \left( 1 + \frac{P \cdot \frac{\ln{n^\alpha}}{c} }{N + P \cdot \frac{n^{\beta (1+\frac{1}{\gamma})-\frac{1}{\gamma}}}{\lambda ^{\frac{1}{\gamma}}(1+\frac{1}{\gamma})}} \right)\cdot
\end{align}

By choosing $\beta=f$, the primary network is protected the the secondary throughput factor is,
\begin{align}
\plim_{n\rightarrow \infty}\frac{R_{\Sigma , \sf{Sim}}^s (n,n^\alpha)}{R_{\Sigma}(n^\alpha)}=
\left\{
\begin{array}{cc}
 \nonumber 1& 0<f \leq\frac{1}{1+\gamma} \\
 \nonumber 0  & \frac{1}{1+\gamma}<f \leq 1,
\end{array}
\right.
\end{align}
for any $\alpha>0$.

Using the joint optimization strategy,
%
we find the upper bounds for the primary and secondary sum-rates to concentrate as
\begin{align}
R_{\Sigma , \sf{UB}, \sf{Sim}}^p (n,n^ \alpha)&\dotsim \log \left( 1 + \frac{P  {n^\beta} \cdot \frac{\ln n}{c}} {N + P G_0^{s\rightarrow p}} \right)\nonumber \\&\dotsim \beta \log n,
\end{align}
and
\begin{align}
R_{\Sigma , \sf{UB}, \sf{Sim}}^s (n,n^ \alpha)&\dotsim \log \left( 1 + \frac{P \cdot \frac{\ln {n^{ \alpha}}}{c}} {N + P \frac{n^{{\beta}(1+\frac{1}{\gamma})} n^{-\frac{1}{\gamma}}}{\lambda^{\frac{1}{\gamma}}(1+\frac{1}{\gamma})} }\right),
\end{align}
respectively.
Again, by solving the optimization problem \eqref{joint:pusu2} subject to \eqref{joint:pusu1}, one can show that no better result is achieved by applying the joint optimization strategy.
\endproof
\section{Primary and Secondary Downlink}
\label{sec:pdsd}
As the primary and secondary networks are both in downlink mode, there is no co-existence level to be considered.
In this scenario, both primary and secondary base stations transmit to their own user with the highest channel power gain. Thus, the sum-rates of the primary and secondary networks are
\begin{align}
    R_{\Sigma , \sf{Sim}}^p (n,n^\alpha)= \log \left(1 + \frac{P\max_{i=1}^n G_i^p}{N + PG_0^{s \rightarrow p}} \right),
   \label{eq:pdsd:1}
\end{align}
and
\begin{align}
   R_{\Sigma , \sf{Sim}}^s (n,n^\alpha) = \log \left(1 + \frac{P\max_{j=1}^{n^\alpha} G_j^s}{N + PG_0^{p \rightarrow s}} \right),
   \label{eq:pdsd:1}
\end{align}
respectively.
Using Lemma 11
\begin{align}
  R_{\Sigma , \sf{Sim}}^p (n,n^\alpha) \dotsim \log \left(1 + \frac{P \cdot \frac{\ln {n}}{c}}{N + P G_0^{s\rightarrow p}} \right)\dotsim \log{\log{n}},
\end{align}
and
\begin{align}
  R_{\Sigma , \sf{Sim}}^s (n,n^\alpha)\dotsim \log \left(1 + \frac{P \cdot \frac{\ln {n^\alpha}}{c}}{N+P G_0^{p\rightarrow s}} \right)\dotsim \log{\log{n}}.
\end{align}

Therefore,
\begin{align}
\plim_{n\rightarrow \infty}\frac{R_{\Sigma , \sf{Sim}}^p (n,n^\alpha)}{R_{\Sigma}(n)} =1 \geq f,
\end{align}
for any $0<f\leq 1$ and
\begin{align}
\plim_{n\rightarrow \infty}\frac{R_{\Sigma , \sf{Sim}}^s (n,n^\alpha)}{R_{\Sigma}(n^\alpha)} =1.
\end{align}
As a result, without any restriction on the value of $\alpha>0$, for any given $0<f \leq 1$, the primary network is always protected, while the secondary achieves a positive throughput factor.
\section{Comparison to TD }
\label{sec:Compare}
The results obtained in the first three uplink-downlink scenarios are valid in case of any arbitrary CDF, $F_G(g)$, which satisfies the two mentioned properties, whereas in the last scenario, only the exponential tail property is required.
As shown in Appendix A, channel gains with either Rayleigh, Rician or Nakagami-m distributions have both properties for some parameters $c>0$, $\lambda>0$ and $\gamma>0$.


 In this section, we compare the maximum secondary throughput factor achievable under simultaneous transmission to that achievable in TD. As it was shown earlier, the asymptotic secondary throughput factor in TD is equal to $1-f$ in all of the uplink-downlink scenarios.

\subsection{Primary and Secondary Uplink:}
In this scenario, generally with asymmetric co-existence when $\alpha> {1+\gamma}$, for $\frac{1+{\gamma \alpha}}{\alpha -\gamma + {\gamma \alpha }}< f\leq 1$ simultaneous transmission outperforms TD.

In case of symmetric co-existence, when $\gamma \geq 1$ for
\begin{align}
\left\{
\begin{array}{ll}
  \frac{\frac{\alpha {\gamma}^2}{\gamma +1}-1}{\gamma \alpha -1 -\gamma} < f \leq 1, & \alpha > 1+{\gamma}\\
 0<f< \frac{\gamma \alpha -\frac{\gamma}{\gamma +1}}{\gamma \alpha +\alpha -\gamma}, & 0<\alpha <\frac{1}{1+\gamma},
 \end{array}
 \right.
\end{align}
the secondary network achieves a greater sum-rate compared to TD.

  As shown in Appendix A, we obtain $\gamma=1$ when channel and interference gains have Rayleigh or Rician distributions. Therefore, in case of Rayleigh or Rician fading, when $\alpha>2$ symmetric co-existence
outperforms TD for $1/2 < f \leq 1$, and when  $0<\alpha<1/2$ symmetric co-existence outperforms TD for $0<f< 1/2$.
Furthermore, with asymmetric co-existence, secondary sum-rate is higher than that in TD for $(1+\alpha)/(2\alpha-1) < f\leq 1$ when $\alpha> 2$.

\begin{figure}[t]
\begin{center}
\epsfig{keepaspectratio = true, width = 3.8 in, figure = 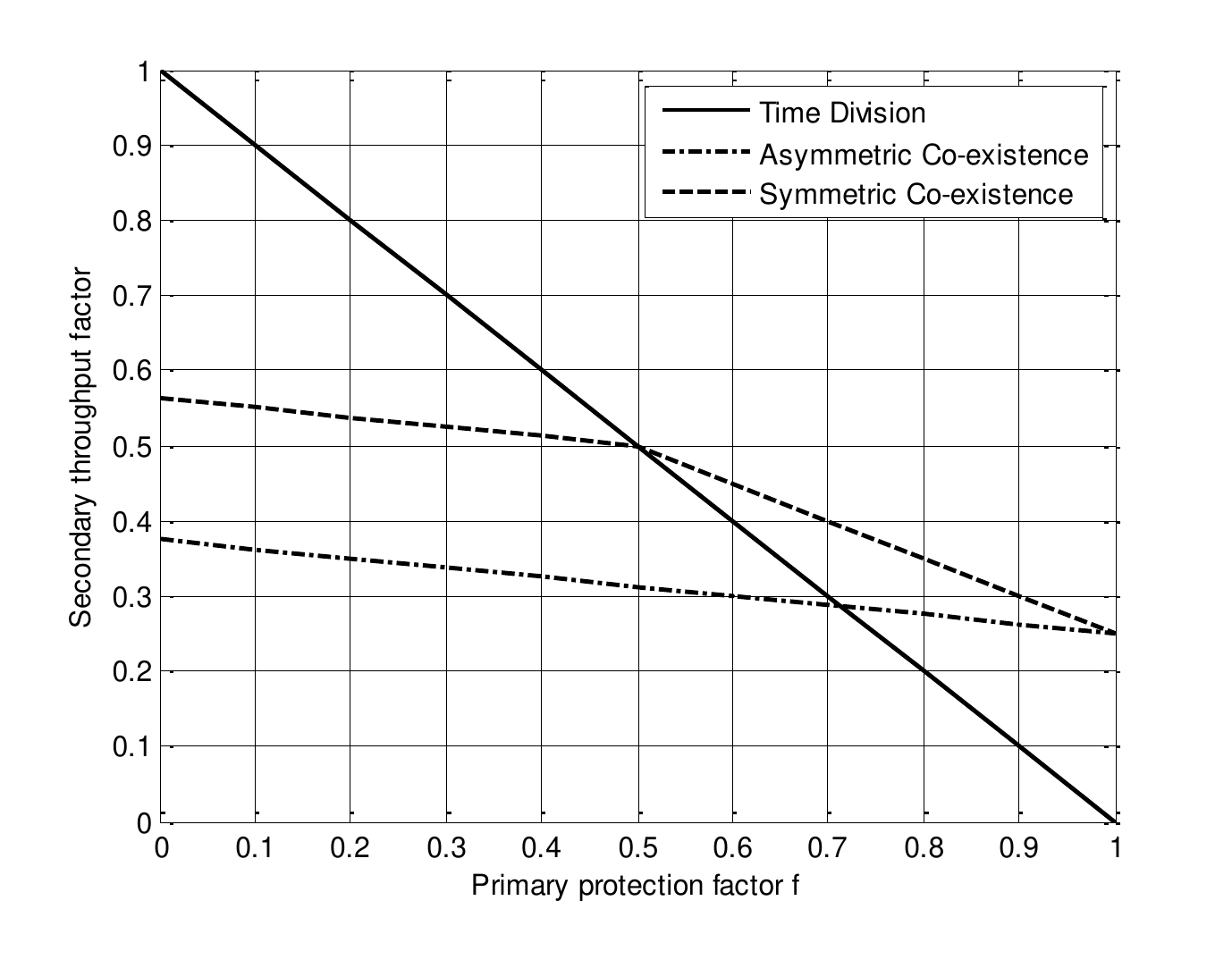}
\end{center}
\caption{Asymptotic result for maximum secondary throughput factor
versus primary protection factor $f$ for $\alpha = 4$ and when both networks are in uplink mode. Rayleigh or Rician distribution is assumed.
Symmetric (resp. asymmetric) co-existence is better than TD
when $1/2< f \leq 1$ (resp. $5/7< f \leq 1$).}
\label{fig:pusu}
\end{figure}

The tradeoff between the secondary throughput factor and the primary protection factor $f$ is illustrated in \Fig \ref{fig:pusu} for $\gamma=1$ and $\alpha=4$.
Surprisingly, when $f=1$, for both symmetric and asymmetric levels, the secondary network can provide positive
sum-rates.

\begin{figure}[t]
\begin{center}
\epsfig{keepaspectratio = true, width = 3.8 in, figure = 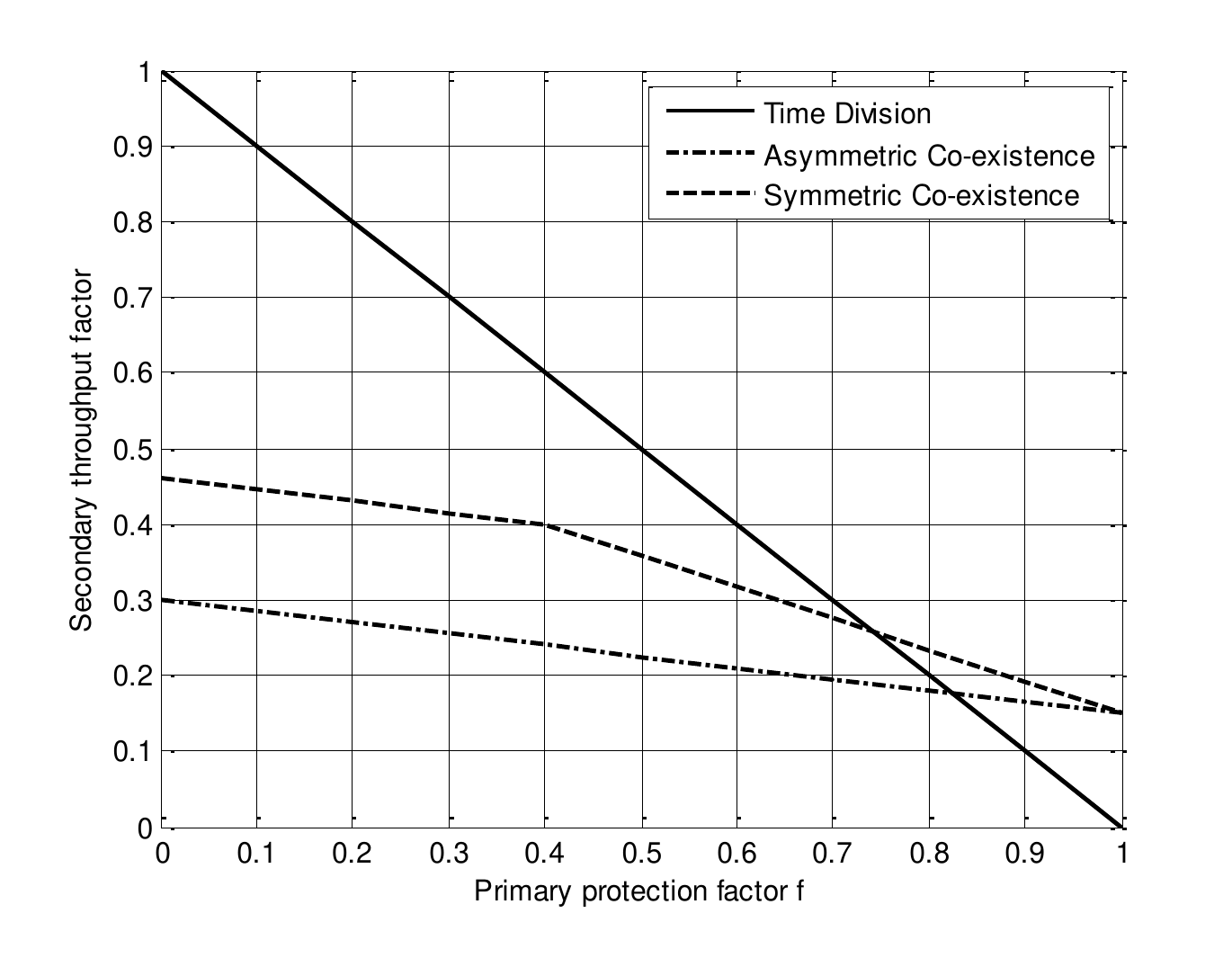}
\end{center}
\caption{Coefficients of sum-rate scalings for secondary
versus primary protection factor $f$ for $\alpha = 4$, when channel and interference gains have Nakagami-m distribution with $m=3/2$ and primary and secondary networks both operate in uplink mode.
Symmetric (resp. asymmetric) co-existence is better than TD
when $ 26/35< f \leq1$ (resp. $ 14/17< f \leq 1$).}
\label{fig:pusu2}
\end{figure}

\Fig \ref{fig:pusu2} shows the secondary throughput factor versus $f$ under simultaneous transmission when channel and interference gains have Nakagami-m distribution with $m=3/2$ and $\alpha=4$. At the asymmetric co-existence level, for $14/17< f\leq 1$, simultaneous transmission is better than TD and at the symmetric co-existence level, simultaneous transmission outperforms TD for $26/35< f \leq 1$.

\subsection{Primary Downlink and Secondary Uplink:} In this scenario, simultaneous transmission asymptotically outperforms TD for $\frac{\gamma}{1+\gamma}< f \leq 1$.
Particularly, in case of Rayleigh or Rician distributions with $\gamma=1$, for $1/2 < f \leq 1$, simultaneous transmission  asymptotically outperforms TD.
When the channel and interference gains have the Nakagami-m distribution with $\gamma =m$, for $\frac{m}{1+m}< f \leq 1$, simultaneous transmission results in a higher sum-rate for the secondary network compared to TD. Decreasing $m$ results in expansion of the interval of $f$ for which a better performance is achievable.
 Since $m$ is reversely proportional to the variance of the Nakagami-m distribution, an increase in the variance improves the range of $f$ for which simultaneous transmission outperforms TD.

\subsection{Primary Uplink and Secondary Downlink:} In simultaneous transmission, for $0<f\leq \frac{1}{1+\gamma}$, the maximum secondary throughput factor is $1$ which is greater than that in TD. In case of Rayleigh and Rician distributions, for $0<f \leq 1/2$, simultaneous transmission results in a higher secondary throughput factor than TD. In case of Nakagami-m distribution, for $0<f\leq \frac{1}{1+m}$, we obtain better results over TD.
Also, smaller $m$ increases the interval of $f$ for which a higher sum-rate is achieved.

\subsection{Primary and Secondary Downlink:} In this scenario, under simultaneous transmission, the secondary network can achieve maximum secondary throughput factor of $1$. Comparing to the secondary throughput factor in TD, which is $1-f$, one can conclude that simultaneous
transmission asymptotically outperforms TD for any $0<f\leq 1$. Thus, when both networks operate in downlink mode, neither is asymptotically impeded by the other. This result is achievable regardless of the value of $\gamma>0$ and therefore we obtain the same result for all of the three distributions.

 Tables \ref{t:results2} and \ref{t:results} summarize the range of protection factor $f$ for which enhanced results over TD are achieved. Table \ref{t:results2} shows the ranges of $f$ for every $\gamma>0$ in the asymmetric co-existence case and the ranges of $f$ when $\gamma\geq 1$ in the symmetric co-existence. The results for the symmetric co-existence case can be applied to the Nakagami-m fading when $m\geq 1$.

 In the primary and secondary uplink scenario with symmetric co-existence, when $\gamma<1$ higher asymptotic sum-rates
  are achieved over TD for some values of $f$. In this case, by comparing the secondary throughput factor derived in Section
   \ref{sec:pusu} and the secondary throughput factor in TD, one can find the ranges of $f$ for which simultaneous transmission
    outperforms TD.
%
%

\begin{table}
\center
 \caption{Ranges of $f$ that result in higher secondary sum-rate over TD in the primary and secondary uplink scenario. For Nakagami-m distribution, $\gamma =m$. For Rayleigh and Rician distributions, $\gamma=1$. }
 \label{t:results2}
\begin{tabular}{ll}
\toprule
Asymmetric & $ \frac{1+{\gamma \alpha}}{\alpha -\gamma + {\gamma \alpha }} < f \leq 1$ ,$\quad$ $\alpha> {1+\gamma}$\\

Symmetric
& $ \frac{\frac{\alpha {\gamma}^2}{\gamma +1}-1}{\gamma \alpha -1 -\gamma} < f \leq 1$, $\quad$ \,$\alpha> 1+{\gamma}$, $\gamma\geq 1$
\\ & $0<f< \frac{\gamma \alpha -\frac{\gamma}{\gamma +1}}{\gamma \alpha +\alpha -\gamma}$, $\quad$ $0<\alpha <\frac{1}{1+\gamma}$, $\gamma \geq 1$ \\
\bottomrule
\end{tabular}
\end{table}

%

\begin{table}
\center
 \caption{Range of $f$ that results in secondary throughput enhancements in the last three scenarios.}
 \label{t:results}
\begin{tabular}{ccc}
\toprule
Primary Network & Secondary Network& Range of $f$   \\
\hline
down & up   & $\frac{\gamma}{1+\gamma} \leq f \leq 1$ \\
up   & down & $0 < f \leq \frac{1}{1+\gamma}$ \\
down & down & $0<f \leq 1$ \\
\bottomrule
\end{tabular}
\end{table}

\section{Value of the Threshold}
\label{sec:Thresholds}
As demonstrated earlier, in the first three scenarios, the best enhanced results are achievable using the least interference strategy in uplink transmission. Thus, such network's scheduler does not require knowledge of the network's own channel power gains in order to activate the users. Therefore, activation of users can be performed in a more distributed way, i.e., it can be performed by a simple comparison between the interference power gain of each user and a certain threshold, followed by activating only the users whose interference power gains are less than the threshold.

Thus, in order to deploy the least interference strategy, only a user's own interference power gain to the opposite network's receiver is required which is easily obtained and is the same as the power gain between the opposite network's transmitter and the user.

If we let $F_G(g)$ be the CDF of the interference power gains, from the discussions in Appendix B, using $F^{-1}_G(\frac{f(n)}{n+1})$ as a threshold gives the best achievable secondary sum-rates with high probability as $n\rightarrow \infty$.
  Therefore, $F^{-1}_G(n^{\beta -1})$ and $F^{-1}_G(n^{{\bar \alpha} -\alpha})$ can be used as thresholds for the primary and secondary networks, respectively. Since the optimal values of $\bar \alpha$ and $\beta$, that achieve the maximum secondary sum-rates in each uplink-downlink scenario are known, the thresholds can be easily derived for any large $n$.


\section{Simulation Results}
\label{sec:Simulation}
Using Lemma 11 and Proposition 12, we have derived asymptotic sum-rates for the primary and secondary networks under simultaneous transmission.
 In this section, in order to demonstrate the accuracy of Lemma 11 and Proposition 12, we provide simulation results.

 We assume that the channel and interference gains are Rayleigh distributed and users are scheduled based on the least interference strategy. We also take $P=10$ dB and $N=0$ dB.

 When the primary and secondary networks are both in uplink mode, for $\beta =0.5$, the sum-rate of the secondary network is
 \begin{align}
R_{\Sigma , \sf{Sim}}^s (n,n^ \alpha) &= \log \left( 1 + \frac{P\sum_{\ell=1}^{n^{\bar{\alpha}}} G^s_{j_\ell}}
                               {N + P \sum_{i=1}^{n^{\beta}} G^{p\rightarrow s}_{i:n}} \right)
                               \label{sim-pusu}
                               \\ \label{sim-pusu1}
                               &\dotsim \log \left( 1+\frac{n^{\bar \alpha}}{\frac{N}{P}+\frac{1}{2}}\right)
                               \\ \label{sim-pusu2} &\dotsim {\bar{\alpha}} \log n.
\end{align}

\begin{figure}[t]
\begin{center}
\epsfig{keepaspectratio = true, width = 3.7 in, figure = 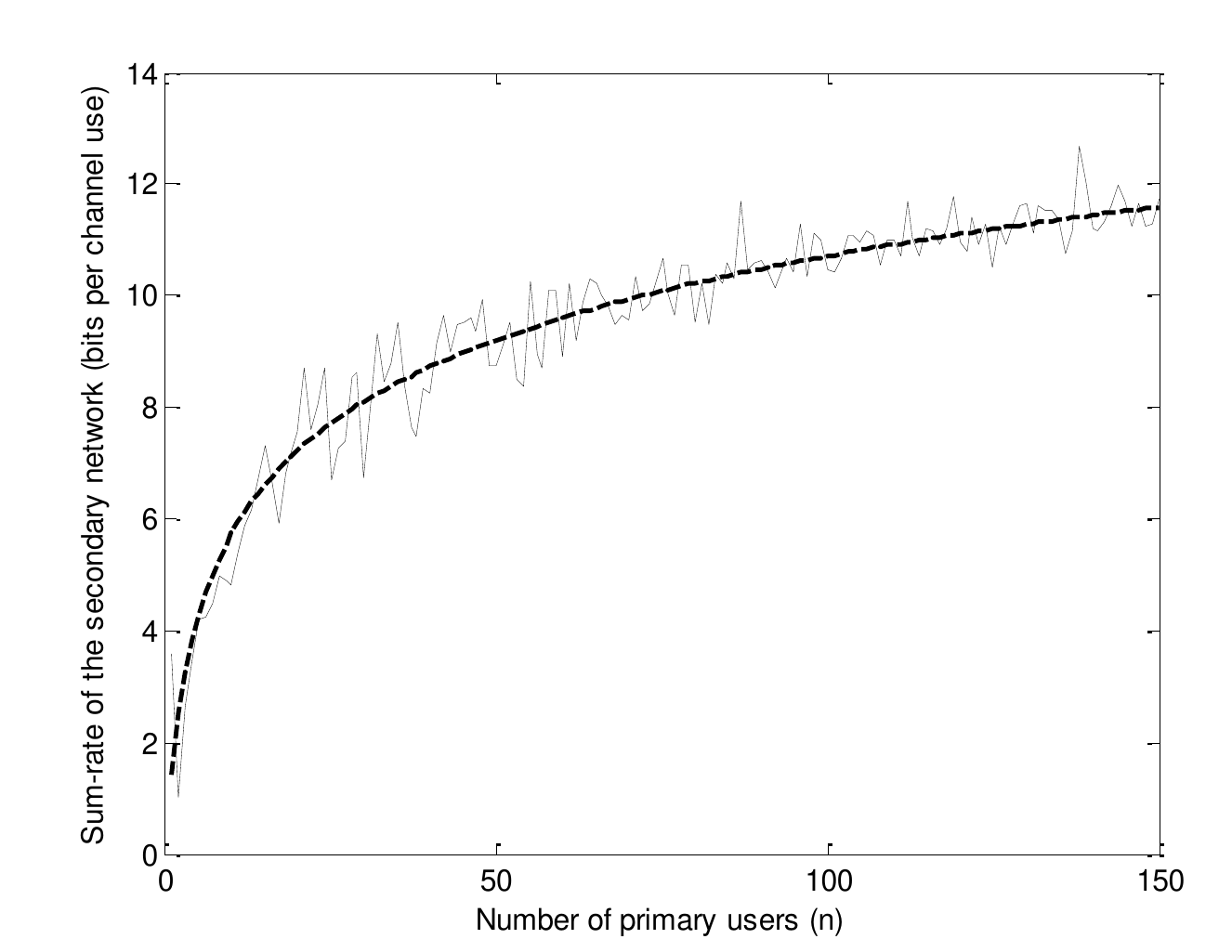}
\end{center}
\caption{The actual (solid) and intermediate (dashed) sum-rates of the secondary network when primary and secondary are both in uplink mode.}
\label{fig:pusu-s-sim}
\end{figure}

We call \eqref{sim-pusu1} an intermediate secondary sum-rate compared to the asymptotic sum-rate in \eqref{sim-pusu2}.
The actual sum-rate, \eqref{sim-pusu}, and the intermediate sum-rate are plotted for $\alpha =3$ and $\bar \alpha =3/2$ in \Fig \ref{fig:pusu-s-sim}.

\begin{figure}[t]
\begin{center}
\epsfig{keepaspectratio = true, width = 3.7 in, figure = 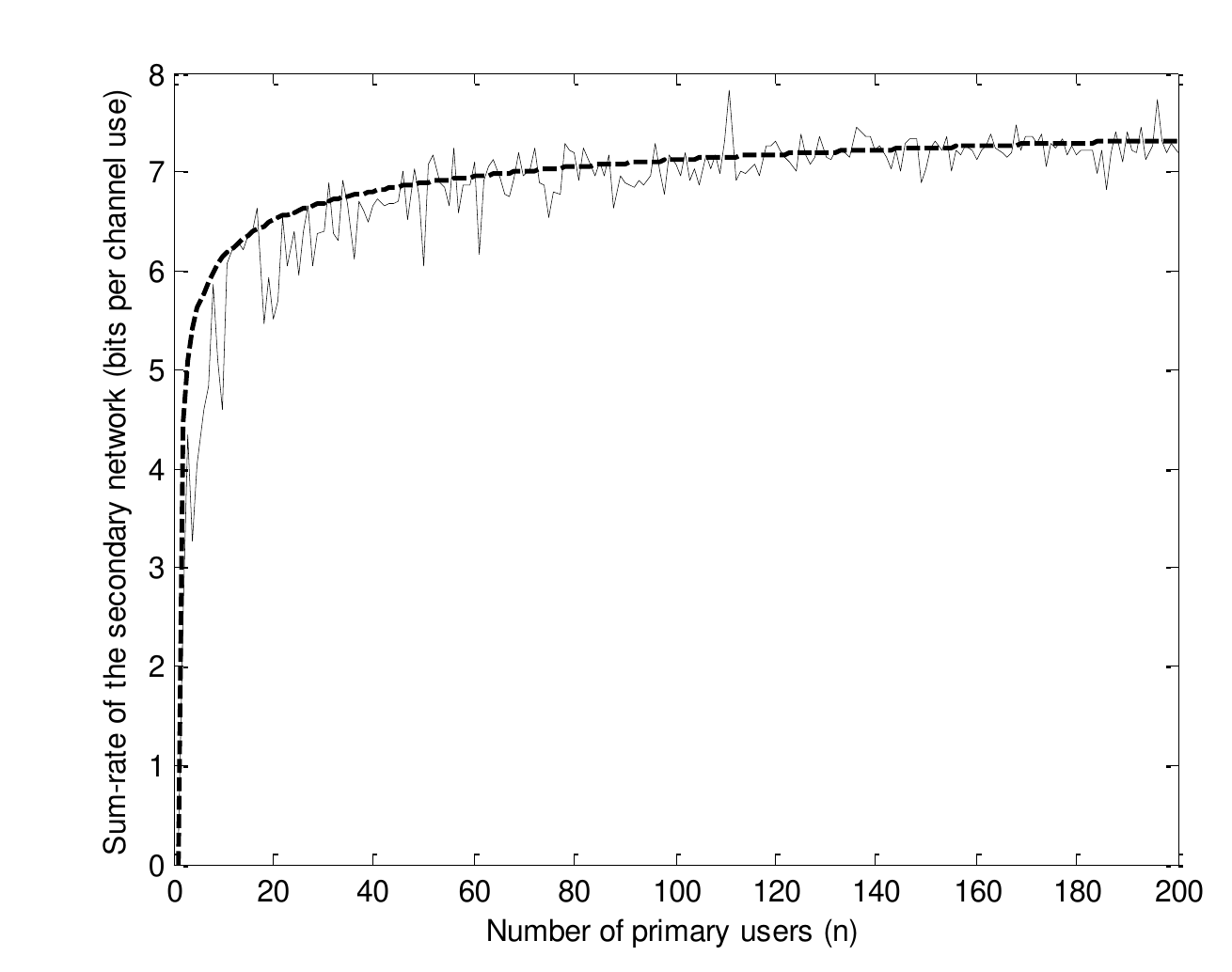}
\end{center}
\caption{The actual (solid) and intermediate (dashed) sum-rate of secondary network when primary is in uplink and secondary is in downlink mode.}
\label{fig:pusd-s-sim}
\end{figure}

When the primary is in uplink and the secondary is in downlink mode, for $\beta=0.5$, the sum-rate of the secondary network is
\begin{align}
  R_{\Sigma , \sf{Sim}}^s (n,n^\alpha) &= \log \left( 1 + \frac{P\max_{j=1}^{n^{\alpha}} G_j^s}{N + P \sum_{i=1}^{n^\beta} G_{i:n}^{p \rightarrow s}} \right) \label{sim-pusd} \\ & \dotsim \log \left( 1+\frac{\ln {n^\alpha}}{\frac{N}{P}+\frac{1}{2}} \right) \label{sim-pusd1} .
\end{align}
The actual, \eqref{sim-pusd}, and the intermediate, \eqref{sim-pusd1}, secondary sum-rates are plotted for $\alpha=3$ in \Fig \ref{fig:pusd-s-sim}.

\begin{figure}[t]
\begin{center}
\epsfig{keepaspectratio = true, width = 3.7 in, figure = 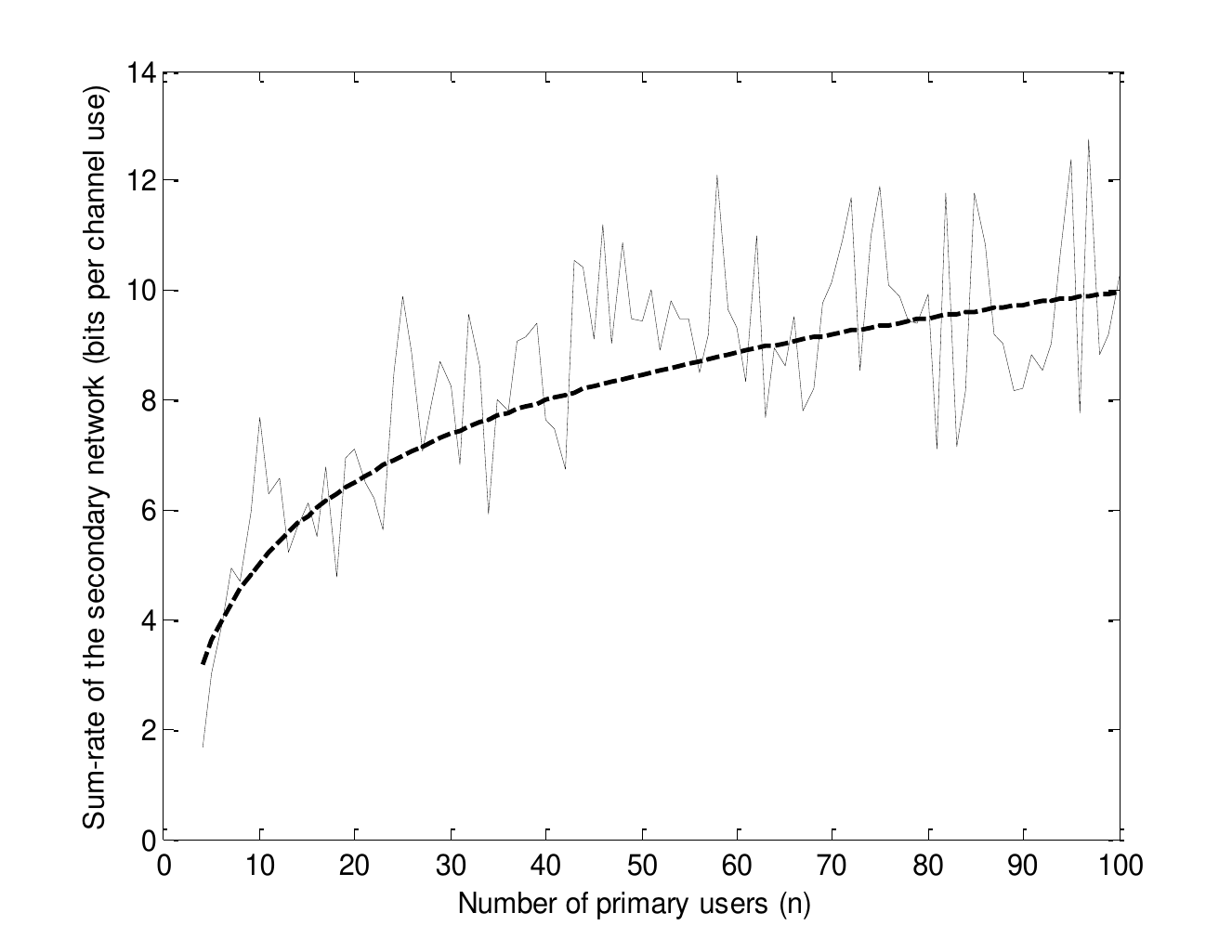}
\end{center}
\caption{The actual (solid) and intermediate (dashed) sum-rate of secondary network when primary is in downlink and secondary is  in uplink mode.}
\label{fig:pdsu-s-sim}
\end{figure}

When the primary is in downlink and the secondary is in uplink mode, the sum-rate of the secondary network is
\begin{align}
  R_{\Sigma , \sf{Sim}}^s (n,n^\alpha) &= \log \left( 1 + \frac{P \sum_{\ell=1}^{n^{\bar{\alpha}}} G^s_{j_\ell}}
                               {N + P G_0^{p\rightarrow s}} \right) \label{sim:pdsu1}\\ & \dotsim \log \left(1+\frac{n^{\bar{\alpha}}}{\frac{N}{P}+ 1} \right)\label{sim:pdsu2}.
\end{align}
The actual, \eqref{sim:pdsu1}, and the intermediate, \eqref{sim:pdsu2}, sum-rates are plotted for $\alpha=3$ and ${\bar{\alpha}}=3/2$ in \Fig \ref{fig:pdsu-s-sim}.

\begin{figure}[t]
\begin{center}
\epsfig{keepaspectratio = true, width = 3.7 in, figure = 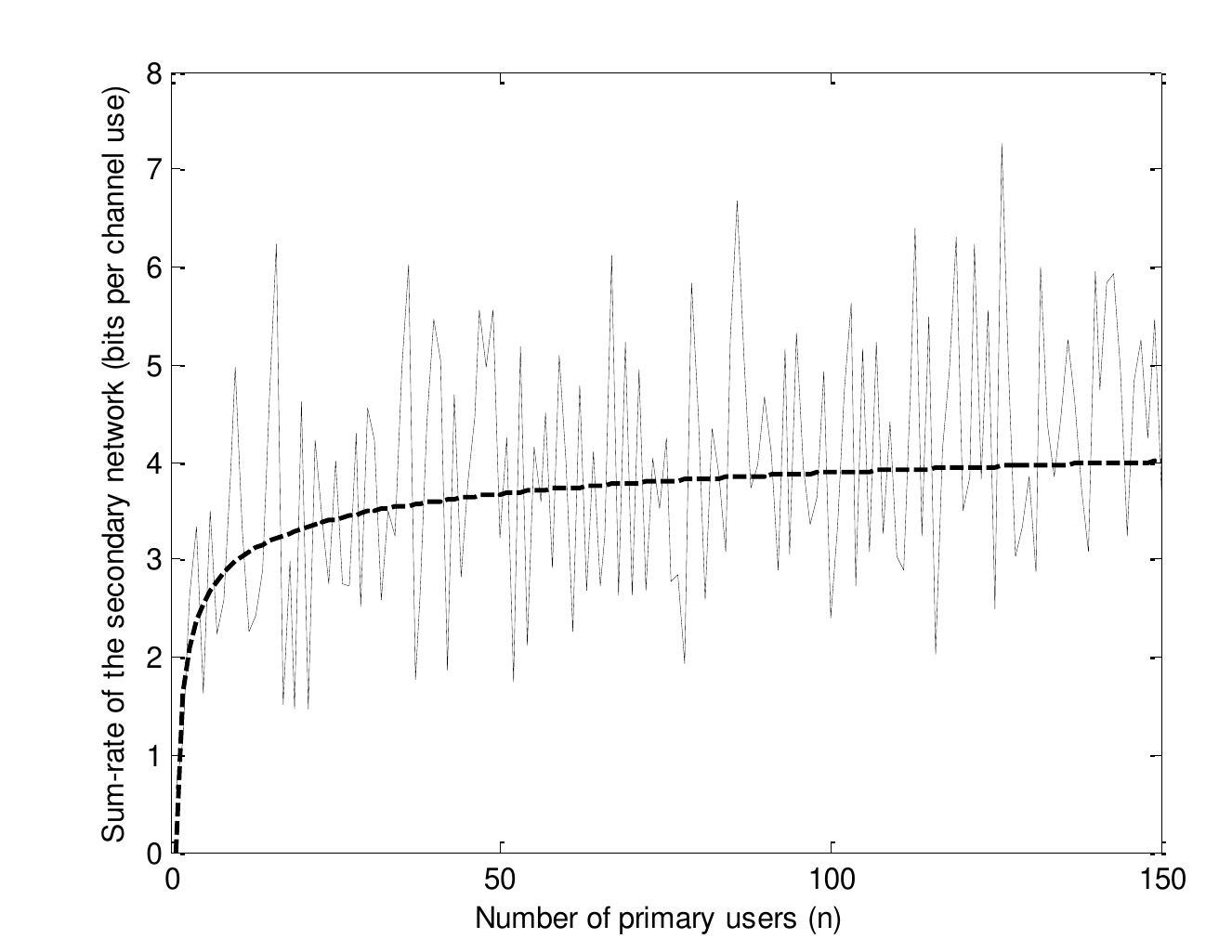}
\end{center}
\caption{The actual (solid) and intermediate (dashed) sum-rate of secondary network when primary and secondary are both in downlink mode.}
\label{fig:pdsd-s-sim}
\end{figure}

When the primary and secondary networks are both in downlink mode, the sum-rate of the secondary network is
\begin{align}
   R_{\Sigma , \sf{Sim}}^s (n,n^\alpha) &= \log \left(1 + \frac{P\max_{j=1}^{n^\alpha} G_j^s}{N + PG_0^{p \rightarrow s}} \right)
   \label{sim-pdsd}\\ & \dotsim \log \left(1+\frac{\ln{n^\alpha}}{\frac{N}{P}+1}  \right)\label{sim-pdsd1}.
\end{align}
The actual, \eqref{sim-pdsd}, and the intermediate, \eqref{sim-pdsd1}, sum-rates are plotted for $\alpha=3$ in \Fig \ref{fig:pdsd-s-sim}.

 As the figures demonstrate, the intermediate sum-rate tracks the actual sum-rate as $n$ increases.
\section{Conclusion}
\label{sec:Conclusion}
In this paper, we have examined the co-existence of primary and secondary point-to-multipoint networks. The networks simultaneously share the same spectrum in a fading environment where the CDF of the interference and channel power gains possess two general properties. We have shown that, at either of the symmetric or asymmetric co-existence levels, a higher asymptotic sum-rate is achievable for the secondary network over TD, while the primary is protected by factor $f$. These enhanced results can be obtained when user scheduling in uplink mode is based only on the least interference gains. In other words, no additional throughput enhancement is made by using the joint optimization strategy.

From a practical standpoint, in uplink transmission, each network can simply compare the interference gains between its users and the opposite network's receiver with a certain threshold and only activate users that generate interferences less than that threshold.
 In downlink mode however, the base stations transmit to their network's strongest user.

 This paper serves as a first step in enhancement of the secondary network's sum-rate in the underlay approach. An extension to this work is to include path loss in the system model. In this case, channel and interference power gains will be products of the corresponding pathloss and the multipath fading gains. Intuitively, this would change the variance of the overall gains and one would expect further improvement with simultaneous transmission compared to TD.


\appendices
\section{}
Here, we show that if the random variable $X$ has either of the three different distributions (Rayleigh, Rician or Nakagami-m) with ${\sf{E}[X^2]}=2$, then the random variable $G={X^2}/2$ has an exponential tail and also satisfies \eqref{proposition}.
Consequently, we will find the parameters $c$, $\gamma$ and $\lambda$ for each distribution.

 Since the Rician distribution reduces to Rayleigh for $K=0$, we find the parameters only for Rician and Nakagami-m fading distributions.\\

 {\em Rician Distribution:}
 Let $X$ be a Rician distributed random variable with $K$-factor and ${\sf{E}[X^2]}=2$, then $G={X^2}/2$ has unit mean and its pdf is \cite{Rician}:
 \begin{align}
 f_G(g)= {(1+K)e^{-K}}\cdot e^{-{(K+1)}g}\cdot I_0(2\sqrt{{K(K+1)g}}),
 \label{app1}
 \end{align}
  where $I_0(.)$ is a modified Bessel function. Then,
 \begin{align}
 \lim _{g\rightarrow\infty}\frac {\ln{f_G(g)}}{g} &= \lim _{g\rightarrow\infty}\frac{\ln{(1+K)e^{-K}}}{g}-(K+1)\\ \nonumber &+\lim _{g\rightarrow\infty}\frac{\ln{I_0(2\sqrt{K(K+1)g})}}{g} \cdot
 \end{align}
As one of the properties of the Bessel function, we have $I_0(\sqrt{g})\sim \frac{e^{\sqrt{g}}}{\sqrt{2\pi \sqrt{g}}}$, when $g \rightarrow \infty$. Then,
\begin{align}
 \lim _{g\rightarrow\infty}\frac {\ln{f_G(g)}}{g}= -(K+1) \cdot
 \end{align}
Thus, $G$ has an exponential tail with parameter $c=(K+1)$.\\

 On the other hand, when $g\rightarrow 0$ we have
\begin{align}
\nonumber I_0(2\sqrt{{K(K+1)g}})& = 1+\frac{4K(K+1)g}{2^2}+\ldots \\ & =1+O(g),
 \end{align}
 and
 \begin{align}
\nonumber e^{-(K+1)g}&=1-(K+1)g+\frac{(K+1)^2 g^2}{2}+\ldots \\ &= 1+O(g)\cdot
 \end{align}

Substituting in \eqref{app1}, we have $f_G(g)= (1+K) e^{-K} [1+O(g)]$. Thus, $F_G (g)=(1+K)e^{-K} g+ O(g^2)$ for $g\rightarrow 0$.
Thus, for Rician distribution with $K$-factor, we obtain the parameters $\lambda = (1+K)e^{-K}$ and $\gamma = 1$.\\

Since Rayleigh distribution is a special case of Rician distribution when $K=0$, if $X$ is Rayleigh distributed with unit mean, then $G$ will have an exponential tail with parameter $c=1$ and for $g\rightarrow 0$ its CDF is in the order of $\gamma=1$ with $\lambda=1$.\\

{\em Nakagami-m Distribution:}
Let $X$ have a Nakagami-m distribution and ${\sf{E}[X^2]}=2$. Then $G={X^2}/2$ will have unit mean and its pdf will be \cite{Nakagami}:
 \begin{align}
 f_G(g)=\frac{1}{\Gamma(m)}{m}^m g^{m-1}e^{-m g},
 \end{align}
 where $m=\frac{{{{\sf{E}} [G]}^2}}{{{\sf{VAR}}[G]}}$.
 Then,
 \begin{align}
 \lim _{g\rightarrow\infty}\frac {\ln{f_G(g)}}{g} &= \lim _{g\rightarrow\infty}\frac{\ln [{\frac{m^m}{\Gamma(m)}g^{m-1}e^{-m g}}]}{g}\\ \nonumber
 &= \lim _{g\rightarrow\infty}\frac{\ln{\frac{m^m}{\Gamma(m)}}}{g}+\lim _{g\rightarrow\infty}\frac{(m-1)\ln g}{g}-m \\ \nonumber
  &= -m \cdot
 \end{align}
Thus, $G$ has an exponential tail with parameter $c=m$.

 On the other hand, when $g\rightarrow 0$
\begin{align}
\nonumber e^{-m g} = 1-m g+ \frac{m^2 g^2}{2} + \ldots = 1+O(g) \cdot
 \end{align}
Then, $f_G(g)= \frac{m^m}{\Gamma(m)}(g^{m-1} +O(g^m))$ and $F_G (g)= \frac{m^{m-1}}{\Gamma(m)}{g^{m}} +O(g^{m+1})$ for $g \rightarrow 0$.
Thus, for the Nakagami-m distribution we obtain $\lambda = \frac{m^{m-1}}{\Gamma(m)}$ and $\gamma = m$.

\section{Proof of Lemma 11 and Proposition 12}

\new{\bf Proof of Lemma 11:}
 Let $X_1, X_2, \ldots, X_n$ be i.i.d. random variables with pdf $f_X(x)$. Let the random variables have an exponential tail with parameter $c>0$, i.e.,  $\lim _{x\rightarrow\infty}\frac {\ln{f_{X}{(x)}}}{x} = -c $.
 Thus, for every real $\epsilon _0> 0$, there exists a real $s_0>0$ such that $\left |\frac{\ln {f_{X}(x)}}{x}+c \right|\leq \epsilon _0 $ whenever $x>s_0 \cdot$ Thus for $x>s_0$
\begin{align}
e^{-(c+\epsilon_0)x}\leq f_{X}(x)\leq e^ {-(c-\epsilon_0)x}.
\label{app2}
\end{align}
Integrating \eqref{app2} from $t>s_0$ to infinity, we will have
\begin{align}
1-\frac{e^{-(c-\epsilon_0)t}}{c-\epsilon_0}\leq F_{X}(t)\leq 1-\frac{e^{-(c+\epsilon_0)t}}{c+\epsilon_0}
\end{align}
 for every $t>s_0$, where $ F_{X}(t)$ is the CDF of random variables $X_1, X_2, \ldots, X_n$.

Define $X_{n:n}=\max_{i=1}^n X_i$. Then from \cite{Arnold}, the CDF of $X_{n:n}$ will be $F_{X_{n:n}}(t)=[F_{X}(t)]^n$. Thus, for every $t>s_0$
\begin{align}
\left[1-\frac{e^{-(c-\epsilon_0)t}}{c-\epsilon_0}\right]^n \leq F_{X_{n:n}}(t)\leq \left[1-\frac{e^{-(c+\epsilon_0)t}}{c+\epsilon_0}\right]^n.
\label{eq1}
 \end{align}
Replacing $t=\left(\frac{1-\epsilon_0}{c+\epsilon_0}\cdot \ln n \right)>s_0$ in \eqref{eq1}, we get
\begin{align}
F_{X_{n:n}}\left(\frac{1-\epsilon_0}{c+\epsilon_0}\ln n \right)\leq \left[ 1-\frac{1}{(c+\epsilon_0)\cdot n^{1-\epsilon_0}}\right]^n.
\end{align}
Replacing $t=\left(\frac{1+\epsilon_0}{c-\epsilon_0}\ln n \right)>s_0$ in \eqref{eq1}, we get
\begin{align}
F_{X_{n:n}}\left(\frac{1+\epsilon_0}{c-\epsilon_0}\ln n \right)\geq \left[ 1-\frac{1}{(c-\epsilon_0)\cdot n^{1+\epsilon_0}}\right]^n.
\end{align}

 It can be verified that,
 \begin{align}
 \lim_{n\rightarrow\infty} \left[ 1-\frac{1}{(c+\epsilon)\cdot n^{1-\epsilon_0}}\right]^n=0 ,
  \end{align}
  and
   \begin{align}
   \lim_{n\rightarrow\infty} \left[ 1-\frac{1}{(c-\epsilon)\cdot n^{1+\epsilon_0}}\right]^n=1.
   \end{align}
  Thus,
\begin{align}
{\sf Pr}\left[\frac{1-\epsilon_0}{c+\epsilon_0}\cdot\ln n \leq X_{n:n} \leq \frac{1+\epsilon_0}{c-\epsilon_0}\cdot\ln n \right] \rightarrow 1,
\end{align}
as $n\rightarrow\infty$, and
\begin{align}
 {\sf Pr}\left[\left|X_{n:n} -\frac {\ln n}{c}\right| \leq \delta \frac {\ln n}{c}\right]\rightarrow 1,
\end{align}
for all $\delta>0$. Thus, $X_{n:n}$ concentrates as $ \frac {\ln n}{c}$.\\

\new{\bf Lemma 16:} Let $g(x) =\lambda x^\gamma+ O(x^{\gamma +1})$ when $x\rightarrow 0$, where $\lambda$ and $\gamma$ are positive real numbers. Then, we have $g^{-1}(x)= {(\frac{x}{\lambda})}^{\frac{1}{\gamma}}+O(x^{\frac{2}{\gamma}})$, when $x\rightarrow 0$.\\

\new{\bf Proof of Lemma 16:} Let
\begin {align}
y=g(x) =\lambda x^\gamma+ O(x^{\gamma +1}),
\end {align}
as $x\rightarrow 0$. Then,
\begin{align}
\label{eq:lemma2}
{(\frac{y}{\lambda})}^{\frac{1}{\gamma}}=x+ O(x^2),
\end{align}
and
\begin{align}
 y^{\frac{2}{\gamma}}=O(x^2),
\label{lemma16-2}
\end{align}
as $x\rightarrow 0$.
Thus, from \eqref{eq:lemma2} and \eqref{lemma16-2} we will have $x = g^{-1}(y)={(\frac{y}{\lambda})}^{\frac{1}{\gamma}}+O(y^{\frac{2}{\gamma}})$ when $y\rightarrow 0$.\\

\new{\bf Proof of Proposition 12:} Let $X_1, \ldots, X_n$ be \iid random variables with CDF  $F_X(x)= \lambda x^\gamma +O(x^{\gamma+1})$ when $x\rightarrow 0$. We denote its order statistics by
 $X_{1:n}, X_{2:n}, \cdots , X_{n:n}$.
Also, let $U_1, \ldots, U_n$ be \iid uniformly distributed random variables defined in $[0,1]$. Then, from \cite{Arnold} we have
$X_{k:n}\eqdist F^{-1}_X{(U_{k:n})} $, where $\eqdist$ denotes equality in distribution.
Thus,
\begin{align}
S_n = \sum_{r=1}^{f(n)} X_{r:n}\eqdist \sum_{r=1}^{f(n)} F^{-1}_X{(U_{r:n})}\cdot
 \end{align}
In \cite{Arnold}, it is shown that
 \begin{align}
 {\sf E}[U_{f(n):n}]= \frac{f(n)}{n+1},
\end{align}
and
\begin{align}
{\sf VAR}[U_{f(n):n}]=\frac{f(n)\left(n+1-f(n)\right)}{(n+1)^2 (n+2)}\cdot
\end{align}
Let $f(n) \rightarrow \infty$ and $f(n)/n \rightarrow 0$ as $n \rightarrow \infty$, then using Tchebychev inequality we will have
\begin{align}
{\sf Pr}\left[ \left|U_{f(n):n}-{\sf E}[U_{f(n):n}] \right|\geq \epsilon \frac{f(n)}{n+1}\right]&\leq \frac{{\sf VAR}[U_{f(n):n}]}{{\epsilon}^2 {{\sf E}[U_{f(n):n}]}^2}\nonumber \\ & \leq \frac{1}{{\epsilon}^2 f(n)}\cdot
\end{align}
Thus, $U_{f(n):n}\dotsim \frac{f(n)}{n+1}$ and for all $\epsilon > 0$, the event
\begin{align}
S_n &\leq \sum _{r=1}^{n} F^{-1}_X{(U_{r:n})} \1{{U_{r:n}\leq {\frac{f(n)}{n+1}(1+\epsilon)}}}\nonumber \\&= \sum _{r=1}^{n} F^{-1}_X{(U_r)} \1{{U_{r}\leq {\frac{f(n)}{n+1}(1+\epsilon)}}},
\end{align}
holds with probability $1-\frac{1}{{\epsilon}^2 f(n)}$
 and the event
 \begin{align}
S_n &\geq \sum _{r=1}^{n} F^{-1}_X{(U_{r:n})} \1{{U_{r:n}\leq {\frac{f(n)}{n+1}(1-\epsilon)}}}\nonumber \\&= \sum _{r=1}^{n} F^{-1}_X{(U_r)} \1{{U_{r}\leq {\frac{f(n)}{n+1}(1-\epsilon)}}},
\end{align}
holds with probability $1-\frac{1}{{\epsilon}^2 f(n)}\cdot$

Now, define
\begin{align}
Y_r=F^{-1}_X{(U_r)} \1{{U_{r}\leq {\frac{f(n)}{n+1}(1+\epsilon)}}},
\end{align}
and
\begin{align}
W_r= F^{-1}_X{(U_r)} \1{{U_{r}\leq {\frac{f(n)}{n+1}(1-\epsilon)}}}.
\end{align}
 Then
\begin{align}
{\sf Pr} \left[\sum_{r=1}^n {W_r} \leq S_n \leq \sum_{r=1}^n {Y_r}\right]\rightarrow 1
\label{propos4}
\end{align}
as $n\rightarrow \infty$.

On the other hand,
\begin{align}
\label{eq:propos}
{\sf E}[Y_r]=\int _0^{\frac{f(n)}{n+1}(1+\epsilon)}F^{-1}_X{(u_r)}du_r,
\end{align}
 \begin{align}
{\sf E}[{Y_r}^2]= \int _0^{\frac{f(n)}{n+1}(1+\epsilon)}[F^{-1}_X{(u_r)}]^2du_r,
\end{align}
and
\begin{align}
{\sf VAR}[Y_r]=E[{Y_r}^2]-{E[Y_r]}^2\cdot
\end{align}

By Lemma 16, $F^{-1}_X(x)={(\frac{x}{\lambda})}^{\frac{1}{\gamma}}+O(x^{\frac{2}{\gamma}})$ when $x\rightarrow 0$. Substituting in \eqref{eq:propos} we will have
\begin{align}
{\sf E}[Y_r]= a\left(\frac{f(n)}{n+1}\right)^{1+\frac{1}{\gamma}}+O\left(\left(\frac{f(n)}{n+1}\right)^{1+\frac{2}{\gamma}}\right),
\end{align}
where $a=\frac{(1+\epsilon)^{1+\frac{1}{\gamma}}}{\lambda^{\frac{1}{\gamma}}(1+\frac{1}{\gamma})} \cdot$
In addition,
\begin{align}
{\sf VAR}[Y_r]= b\left(\frac{f(n)}{n+1}\right)^{1+\frac{2}{\gamma}}+O\left(\left(\frac{f(n)}{n+1}\right)^{2+\frac{2}{\gamma}}\right),
\end{align}
where $b=\frac{\left(1+\epsilon \right)^{1+\frac{2}{\gamma}}}{\lambda^{\frac{2}{\gamma}}\left(1+\frac{2}{\gamma}\right)}$.
Also, since
\begin{align}
 {\sf E}\left[\sum _{r=1}^{n} Y_r \right]=n {\sf E}[Y_r],
\end{align}
and
\begin{align}
 {\sf VAR}\left[\sum _{r=1}^{n} Y_r \right]=n {\sf VAR}[Y_r],
\end{align}
using Tchebychev inequality we have
\begin{align}
 {\sf Pr}\left[ \left|\sum _{r=1}^{n} Y_r- n {\sf E}[Y_r]\right|  \geq \delta n {\sf E}[Y_r] \right]
&\leq \frac{{\sf VAR}[Y_r]}{n \delta^2 {\sf E}[Y_r]^2}
\leq \frac{1}{4\delta^2 f(n)}\cdot
\end {align}
Hence, for all $\epsilon >0$
\begin{align}
\label{propos2}
\sum _{r=1}^{n} Y_r \dotsim n \cdot\frac{\left(1+\epsilon \right)^{1+\frac{1}{\gamma}}}{\lambda^{\frac{1}{\gamma}}\left(1+\frac{1}{\gamma}\right)} \cdot \left(\frac{f(n)}{n}\right)^{1+\frac{1}{\gamma}}\cdot
\end{align}

Similarly, one can show that
\begin{align}
\label{propos3}
\sum _{r=1}^{n} W_r \dotsim n \cdot\frac{\left(1-\epsilon \right)^{1+\frac{1}{\gamma}}}{\lambda^{\frac{1}{\gamma}}\left(1+\frac{1}{\gamma}\right)} \cdot\left(\frac{f(n)}{n}\right)^{1+\frac{1}{\gamma}},
\end{align}

and, from \eqref{propos4}, \eqref{propos2} and \eqref{propos3} we will have
\begin{align}
S_n \dotsim \frac{n}{\lambda^{\frac{1}{\gamma}}\left(1+\frac{1}{\gamma}\right)} \cdot\left(\frac{f(n)}{n}\right)^{1+\frac{1}{\gamma}}\cdot
\end{align}



\end{document}